\documentclass{birkart04}

\usepackage{amsmath}
\usepackage{amssymb}
\usepackage{amsfonts}
\usepackage{latexsym}
\usepackage{color}

 \normalbaselines

\catcode `\@=11
\@addtoreset{equation}{section}
\def\theequation{\arabic{section}.\arabic{equation}}
\catcode`\@=12

\newtheorem{theo}{Theorem}[section]

\newtheorem{lem}[theo]{Lemma}
\newtheorem{prop}[theo]{Proposition}
\newtheorem{cor}[theo]{Corollary}

\newtheorem{definition}[theo]{Definition}

\newtheorem{example}[theo]{Example}

 \numberwithin{equation}{section}

\newtheorem{remarks}[theo]{Remarks}
\newtheorem{remark}[theo]{Remark}

\newcommand{\betheo}{\begin{theo}$\!\!\!${\bf } }
\newcommand{\entheo}{\end{theo}}

\newcommand{\becor}{\begin{cor}$\!\!\!$  }
\newcommand{\encor}{\end{cor}}

\newcommand{\belem}{\begin{lem}$\!\!\!$  }
\newcommand{\enlem}{\end{lem}}

\newcommand{\beprop}{\begin{prop}}
\newcommand{\enprop}{\end{prop}}

\newcommand{\bedefi}{\begin{definition}$\!\!\!$ \rm }
\newcommand{\findefi}{ \end{definition}}

\newcommand{\beex}{\begin{example}$\!\!\!$ \rm }
\newcommand{\enex}{ \end{example}}

\newcommand{\berem}{\begin{remark}$\!\!\!$ \rm }
\newcommand{\enrem}{ \end{remark}}

\newcommand{\berems}{\begin{remarks}$\!\!\!$ \rm }
\newcommand{\enrems}{ \end{remarks}}

\newcommand{\be}{\begin{equation}}
\newcommand{\en}{\end{equation}}

\newcommand{\bea}{\begin{eqnarray}}
\newcommand{\ena}{\end{eqnarray}}

\newcommand{\beano}{\begin{eqnarray*}}
\newcommand{\enano}{\end{eqnarray*}}

\newcommand{\bee}{\begin{enumerate}}
\newcommand{\ene}{\end{enumerate}}

\newcommand{\bei}{\begin{itemize}}
\newcommand{\eni}{\end{itemize}}

\newcommand{\betab}{\begin{tabular}}
\newcommand{\entab}{\end{tabular}}

\newcommand{\up}{\raisebox{0.7mm}{$\upharpoonright $}}%

\newcommand{\nn}{\nonumber}

\newcommand{\ov}[1]{\overline{#1}}

\newcommand{\mc}{\mathcal}
\newcommand{\mb}{\mathbb}
\newcommand{\RN}{\mb R}
\newcommand{\CN}{\mb C}
\def\NN{{\mathbb N}}
\def\ZN{{\mathbb Z}}

\def\gF{{\mathfrak F}}
\def\A{{\mathcal A}}
\def\B{{\mathcal B}}
\def\C{{\mathcal C}}
\def\D{{\mathcal D}}
\def\E{{\mathcal E}}
\def\F{{\mathcal F}}

\def\H{{\mathcal H}}

\def\J{{\mathcal J}}
\def\K{{\mathcal K}}

\def\O{{\mathcal O}}

\def\cS{{\mathcal S}}

\def\gF{{\mathfrak F}}
\def\h{{\mathfrak H}}
\newcommand{\bPhi}{\boldsymbol\Phi}
\newcommand{\bB }{\mathbf B}

\newcommand{\norm}[2]{\left\| #2 \right\|_{#1}}
\newcommand{\dis}{\displaystyle}

\newcommand{\noi}{\noindent}
\newcommand{\ud}{\,\mathrm{d}}
\renewcommand{\leq}{\leqslant}
\renewcommand{\geq}{\geqslant}

\def\hs{Hilbert space}

\def\dag{\dagger}

\newcommand{\ip}[2]{\langle {#1} |{#2} \rangle}

\def\OL{\relax\ifmmode {\sf L}\else{\textsf L}\fi}
\def\OR{\relax\ifmmode {\sf R}\else{\textsf R}\fi}

\newcommand{\ta}{^\times}

\newcommand{\bdim}{ {\bf Proof. }}
 \newcommand{\edim}{\qed}
\newcommand{\pip}{{\sc pip}-space}
\newcommand{\ipip}{indexed {\sc pip}-space}

 \newcommand{\co}{^{\#}}

\newcommand{\com}{{\scriptstyle\#}}

\newcommand{\s}{\underline}

\begin{document}

\title{Operators on Partial Inner Product Spaces: Towards \\a Spectral Analysis}
\author[Antoine]{Jean-Pierre Antoine}
\address{%
Institut de Recherche en Math\'ematique et  Physique\\
Universit\'e Catholique de Louvain\\
B-1348   Louvain-la-Neuve\\
Belgium}
\email{jean-pierre.antoine@uclouvain.be}

\author[Trapani]{Camillo Trapani}
\address{%
Dipartimento di Matematica e Informatica \\
Universit\`a di Palermo\\
I-90123 Palermo\\
Italy}
\email{camillo.trapani@unipa.it}


\date{\today}

\begin{abstract}
Given a LHS (Lattice of Hilbert spaces) $V_J$ and  a symmetric operator $A$ in $V_J$, in the sense of partial inner product spaces, we define a generalized resolvent for $A$ and study the corresponding  spectral properties. In particular, we examine, with help of the  KLMN theorem, the question of generalized eigenvalues associated to points of the continuous (Hilbertian) spectrum. We give some examples, including so-called frame multipliers.
\end{abstract}

\maketitle

\section{Introduction}
\label{sect-intro}

In physics, rigged \hs s (RHS) are standard tools in  Quantum Mechanics, in particular for reconciling the   convenient bra-ket formalism of Dirac with the mathematically rigorous approach of von Neumann \cite[Chap.7]{pip-book}. In particular, the question of generalized eigenvalues of observables, associated to points of the continuous spectrum, is solved with help of the celebrated Maurin-Gel'fand theorem.

In  a recent paper, Bellomonte et al. \cite{bello-db-trap} have attacked this problem by considering observables as operators in
 ${\mc L}(\D,\D\ta)$, for a suitable RHS $\D \subset \H \subset \D\ta$, where ${\mc L}(\D,\D\ta)$ is the space of all continuous linear maps from $\D$ into $\D\ta$. However, the framework they use in a large part of their paper is in fact a partial inner product space (\pip), more precisely  a LHS (Lattice of \hs s).

 Indeed, the basic ingredient in \cite{bello-db-trap} is that of a family $\gF$ of interspaces between $\D$ and $\D\ta$
  \cite[Sec. 5.4.1]{pip-book}. By interspace, one means a locally convex space $\E[\tau(\E,\E\ta)]$, equipped with the Mackey topology from its conjugate dual, and such that $\D \subset \E \subset \D\ta$, where both embeddings are continuous and have dense range. In addition, one requires that the family  $\gF$ of interspaces be a multiplication framework, that is, (i) $\D\in \gF$; (ii) for every $\E \in \gF$, the conjugate dual $\E\ta$ also belongs to  $\gF$; and (iii) for every pair $\E,\F \in \gF, \, \E \cap\F \in \gF$. Then, if every interspace $\E \in \gF$ (except $\D$ and $D\ta$) is a \hs, as assumed in most of \cite{bello-db-trap}, the resulting structure is a LHS (Lattice of \hs s)   $V_J$ in the sense of \cite{pip-book} and   ${\mc L}(\D,\D\ta) \equiv \mathrm{Op}(V_J)$.

 In view of this fact, we feel the analysis becomes simpler if one uses  the language of \pip s from the beginning.
 Thus we will make a few steps towards  a spectral theory of symmetric operators in a LHS, following in part
   \cite{bello-db-trap}. Our framework will be a LHS $V_J$ and we adopt the definitions and notations of our monograph \cite{pip-book}. For the convenience of the reader, we summarize in the Appendix the salient features of \pip s and operators on them.

   The paper is organized as follows. Section 2 is devoted to the notion of inverse operator
in the \pip\ context, with application to resolvents and, in particular, their analyticity properties. In Section 3, we discuss the various aspects of spectral analysis of \hs\ operators, including the generalized eigenvalues and eigenvectors, in the light of the well-known KLMN theorem. In particular we revisit the notion of  \emph{tight rigging}. Section 4, finally, is devoted to several examples of spectral analysis of rather singular operators. As for notations, the domain of a \hs\ operator $A$ is denoted $D(A)$ and its range by $\mathrm{Ran} (A)$.

\section{Inverses and resolvents}
\label{sec-resolv}

\subsection{Invertible operators}
\label{subsec-invert}

The key ingredient of the spectral theory of operators  is the notion of \emph{resolvent}. For fixing ideas, given    a closed operator $A$ in a
\hs\  $\H$, consider $A-\lambda I : D(A) \to \H$. {Then the resolvent of $A$ is $R_\lambda(A) := (A-\lambda I)^{-1},$ for those $\lambda\in {\mb C}$ for which this inverse exists as an everywhere defined bounded operator in $\H$, that is, $\lambda\in \rho(A)\subset \CN$, the resolvent set of $A$}.
In order to extend this notion to a \pip, we have first to define an appropriate concept of inverse of an operator, and this is nontrivial.

Let $V_J$ be a  LBS/LHS and $A\in \mathrm{Op}(V_J)$. According to \cite[Sec. 3.3.2]{pip-book},
we shall say that a representative $A_{pq}$ is \emph {invertible} if it is bijective, hence it has a continuous inverse $B_{qp} := (A_{pq})^{-1}: V_p \to V_q$ .
  Any successor $A_{p'q'}, q'\leq q, p'\geq p,$ of an invertible representative $A_{pq}$ is injective and has dense range. An invertible representative has in general no predecessors, that is, a representative $A_{p'q'}$ with $q'\geq q, p'\leq p,$.   This does not exclude  the possibility for an invertible operator $ A$ to have a nontrivial null-space. Indeed, $A$ may have a noninjective representative $A_{sr}$, where $r$ is not comparable to $q$, i.e., there may exist a $g\in V_r$ such that $Ag=0$, provided $g\not\in V_q$.
   Note that, if $A_{pq}$ is  invertible, $A^\times_{\ov{q},\ov{p}}$ is  also invertible and
$(A^\times_{\ov{q},\ov{p}})^{-1} =  (A_{pq}^{-1})' : V_{\ov{q}} \to  V_{\ov{p}}$.

Given an operator $A\in \mathrm{Op}(V_J)$, we recall that $(q,p)\in {\sf j}(A)$  means that $A$ has a continuous representative
$A_{pq}: V_{q} \to V_{p}$.
\belem\label{lem-inv}
 Let $V_J$ be a  LBS/LHS and $A\in \mathrm{Op}(V_J)$. Then the following conditions are equivalent:

 (i) $A$ has an invertible representative.

 (ii) There exist an operator $B\in \mathrm{Op}(V_J)$ and two indexes $p,q$ such that $(p,q)\in {\sf j}(A),\,
 (q,p)\in {\sf j}(B)$, and $AB=BA = I$.
\enlem
\bdim
\s{(i) $\Rightarrow$ (ii):}  Let $A_{pq}$ be invertible. Since $(A_{pq})^{-1}: V_p \to V_q$ is continuous, it defines a unique operator $B\in \mathrm{Op}(V_J)$, by
$ B_{qp} = (A_{pq})^{-1}$.  Thus $AB$ and $BA$ are well defined, and $A_{pq}B_{qp}= I_{pp},
B_{qp}A_{pq}= I_{qq}$, that is, by the maximality property of \pip\ operators,  $AB=BA = I$.

 \s{(ii) $\Rightarrow$ (i):} By the assumption, $AB$ and $BA$ are well defined. Since $AB=BA = I$, we may write
$A_{pq}B_{qp}= I_{pp}, \;B_{qp}A_{pq}= I_{qq}$. Then, the first condition implies that $A_{pq}$ is surjective and
 $B_{qp}$ is injective, whereas the second condition implies that $B_{qp}$ is surjective and $A_{pq}$ is injective. Thus they are both bijective, hence boundedly invertible.
\edim
\medskip

We note that an algebraic condition, namely, $AB, BA$ are well defined  and $AB=BA = I$, is \emph{not} sufficient.
 Therefore, on the basis of the previous lemma, we  define invertibility of an \pip\ operator as follows.
\bedefi\label{invertible}
 Given a LBS/LHS, an operator $A\in \mathrm{Op}(V_J)$ is \emph{invertible}   if it has at least one  invertible representative.
\findefi

Of course,  the operator $B$ defined by Lemma \ref{lem-inv} (ii)   will be called an \emph{inverse} of $A$, but it remains to show that it is unique.
   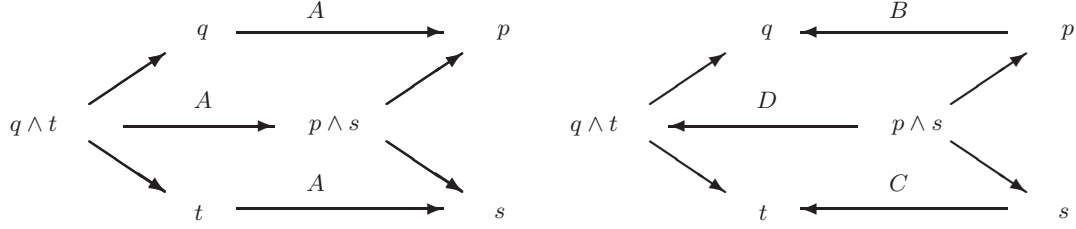
\begin{figure}[t]
\centering \setlength{\unitlength}{0.5cm}
\begin{picture}(8,8)
\put(0,4){
\begin{picture}(8,8) \thicklines
\small
\put(-8.9,0.6){\vector(3,2){2}}
\put(-8.9,-0.4){\vector(3,-2){2}}

\put(-1,0.6){\vector(3,2){2}}
\put(-1,-0.4){\vector(3,-2){2}}

\put(-6,2.5){\makebox(0,0){ $q$}}
\put(-10.5,0){\makebox(0,0){ $q\wedge t$}}
\put(-6,-2.3){\makebox(0,0){$t$}}

\put(-5,2.5){\vector(1,0){5.5}}
\put(-5,-2.2){\vector(1,0){5.5}}
\put(-8,0){\vector(1,0){4}}

\put(-3,3.1){\makebox(0,0){ $A$}}
\put(-6,0.7){\makebox(0,0){ $A$}}
\put(-3,-1.5){\makebox(0,0){ $A$}}

\put(2,2.5){\makebox(0,0){ $p$}}
\put(2,-2.3){\makebox(0,0){$s$}}
\put(-2.5,0){\makebox(0,0){ $p\wedge s$}}

\put(-8.9,0.6){\vector(3,2){2}}
\put(-8.9,-0.4){\vector(3,-2){2}}

\put(-1,0.6){\vector(3,2){2}}
\put(-1,-0.4){\vector(3,-2){2}}
\put(6,0.6){\vector(3,2){2}}
\put(6,-0.4){\vector(3,-2){2}}

\put(14,0.6){\vector(3,2){2}}
\put(14,-0.4){\vector(3,-2){2}}

\put(9,2.5){\makebox(0,0){ $q$}}
\put(4.4,0){\makebox(0,0){ $q\wedge t$}}
\put(9,-2.3){\makebox(0,0){$t$}}

\put(15.5,2.5){\vector(-1,0){5.5}}
\put(15.5,-2.2){\vector(-1,0){5.5}}
\put(11.5,0){\vector(-1,0){5}}

\put(12.5,3.1){\makebox(0,0){ $B$}}
\put(9,0.7){\makebox(0,0){ $D$}}
\put(12.5,-1.5){\makebox(0,0){ $C$}}

\put(17,2.5){\makebox(0,0){ $p$}}
\put(17,-2.3){\makebox(0,0){$s$}}
\put(13,0){\makebox(0,0){ $p\wedge s$}}

\end{picture}
}
\end{picture}
\caption{\label{fig:diagram}Action of the various operators. Each (assaying) space is represented by its index.}

\end{figure}

\beprop\label{prop-inverse}
{Let  $V_J$ be a  LBS/LHS and $A\in \mathrm{Op}(V_J)$   an invertible operator. Then $A$   has a unique inverse $A^{-1}\in \mathrm{Op}(V_J)$.}
\enprop
\bdim
 If $A$ has the invertible representative $A_{pq}$, we know that it has an inverse $B$, such that $AB=BA = I$.
    Suppose  now that $A$ has two invertible representatives $A_{pq}$ and $A_{st}$, that is, they are both bijective and continuous.  In the same way as $(A_{pq})^{-1}$ defines a unique operator  $B\in \mathrm{Op}(V_J)$,  $(A_{st})^{-1}$ defines a unique operator  $C\in \mathrm{Op}(V_J)$
by $ C_{ts} = (A_{st})^{-1}: V_s \to V_t$. Thus we have $C_{ts}A_{st}= I_{tt}$ and $A_{st}C_{ts}= I_{ss}$. Hence
we may write $AC = CA = I$, that is,  $C$ is also an   inverse of $A$.
We claim that $B=C$.
First, $B_{qp}$ and $C_{ts}$ have well-defined restrictions to $V_{p\wedge s}$, namely,
 $B_{q, p\wedge s}$, resp. $C_{t, p\wedge s}$.
 Next,  according to Lemma 3.3.29 of \cite{pip-book}, $(q\wedge t,p\wedge s)$
 belongs to ${\sf j}(A)$  and  $A_{p\wedge s, q\wedge t}$
 is also bijective and continuous.
Hence $A_{p\wedge s, q\wedge t}$ has a continuous inverse
$D_{q\wedge t,p\wedge s} = (A_{p\wedge s, q\wedge t})^{-1} : V_{p\wedge s} \to V_{q\wedge t}$. The latter defines a unique operator
 $D\in \mathrm{Op}(V_J)$, that is another inverse of $A$. One has indeed:
\begin{align*}
 \mathrm{For}\; f\in V_p, \; (ABf)_p &= A_{pq}\,B_{qp}\,f_p = f_p \, ,
\\
 \mathrm{For}\; g\in V_s, \; (ACg)_s &= A_{st}\,C_{ts}\,g_s = g_s \, ,
\\
 \mathrm{For}\;  h\in V_{p\wedge s}, \; (ADh)_{p\wedge s}
&= A_{p\wedge s, q\wedge t}\,D_{q\wedge t,p\wedge s}\,h_{p\wedge s} = h_{p\wedge s}\, .
\end{align*}
We refer to Fig. \ref{fig:diagram} for the action of the various operators $A,B,C,D$.

Clearly,  $A_{p\wedge s, q\wedge t}$ is the restriction of $A_{pq}$ to $V_{q\wedge t}$. In the same way,
 $D_{q\wedge t,p\wedge s} $ is the restriction of $B_{qp}$ to $V_{p\wedge s}$.
Similarly,  $D_{q\wedge t,p\wedge s} $ is also the restriction of $C_{ts}$ to $V_{p\wedge s}$.
Thus $B_{qp}$ and $C_{ts}$ have the same restriction to $V_{p\wedge s}$, \emph{a fortiori} to $V\co$, which implies that $B=C\in \mathrm{Op}(V_J)$. (This supersedes Remark 3.3 of \cite{bello-db-trap}).
\edim

\berems (1)  If $B$ is an inverse of $A$, we have written  $AB = BA = I$, but this requires some qualification. In the case of  an unbounded invertible operator $X$ in a \hs, one has to write $X^{-1}\, X \subset I$, instead of  $X^{-1}\, X = I$, because the l.h.s. has a dense domain, whereas the identity is everywhere defined. But in a \pip, the notion of extension of an operator does not exist, every operator is maximal, by definition (see Sec. A.2). The inverse condition
$B_{qp}A_{pq}= (BA)_{qq} = I_{qq}$ means, first,  that the product $BA$ is well defined, then that it coincides with the identity on $V_q$.
Since a single representative determines a unique operator in $\mathrm{Op}(V_J)$, it follows that $ BA = I$ as \pip\ operators.
The same reasoning applies if one restricts oneself to $V\co$: if $BAf = f $, for every $f\in V\co$, one has again  $ BA = I$.  .

We emphasize that, in general, the product $BA$ may have many more representatives (that is, it can be better behaved) than the operator $A$ itself, because of the maximality axiom. This is precisely the case here.

(2) As a final remark, we may note that the crucial Lemma 3.3.29 of \cite{pip-book} is true for any projective, positive definite \ipip, that is, an \ipip, in which any intersection $V_{p\wedge q} = V_{p} \cap V_{ q}$ carries the projective topology inherited from $V_{p}$ and $V_{ q}$. This is the case when both spaces are Fr\'echet spaces, in particular, for a LHS/LBS.
Thus, whereas  Lemma \ref{lem-inv} holds in general, uniqueness of the inverse  is valid only in the projective case, since the proof of Proposition \ref{prop-inverse} relies on the Lemma 3.3.29 of \cite{pip-book}.
\enrems

\subsection{Regular points}
\label{subsec-regular}

In this section we will extend to \pip s the notion of regular points familiar in \hs\ theory (see \cite[Chap.2]{schm}  or \cite[Chap.8]{weidmann}).

 \bedefi
 A number $\lambda\in \CN$ is called a \emph{$J$-regular} point for $A\in \mathrm{Op}(V_J)$, if there exist  $(q,p)\in  {\sf j}(A)$, with $q\leq p$,  and constants $c_\lambda, d_\lambda$  such that
\be\label{eq:regular}
 c_\lambda \norm{q}{f} \leq \norm{p}{(A-\lambda I)f}\leq   d_\lambda \norm{q}{f}, \; \forall\, f\in V_q \, .
\en
\findefi
Note that the upper bound (which is absent in the \hs\ context) results simply from the fact the the representative
$A_{pq}$ is bounded.
The set of $J$-regular points of $A$ will be denoted by  $\pi^J(A)$.
Clearly, $\lambda\in\pi^J(A)$ if and only if  there exist  $(q,p)\in  {\sf j}(A)$, with $q\leq p$,  such that
$(A-\lambda I)_{pq}$ is injective.
We also denote by $ \pi^{(q,p)}(A)$ the set of $J$-regular points of $A$ for fixed $q,p$.

Actually \eqref{eq:regular} implies that that the inverse of $(A-\lambda I) _{pq}: V_q\to V_p$ is bounded, but it is defined only on $\mathrm{Ran} (A-\lambda I)_{pq}$, which need not be the whole of $V_p$.
For $\lambda\in  \pi^{(q,p)}(A)$, call $d^{(q,p)}_\lambda(A) := \mathrm{dim}\,[\mathrm{Ran} (A-\lambda I)_{pq}] ^\bot$, the orthogonal being taken in $V_p$, the \emph{defect number} of $A$ at $\lambda$ with respect to $p,q$.

Let $(q,p)\in  {\sf j}(A)$. Then the following relations are immediate:
\bei
\item if $p\leq p'$, then $d^{(q,p)}_\lambda(A)\leq d^{(q,p')}_\lambda(A)$.
\item if $p'\leq p $ and $\mathrm{Ran} (A-\lambda I)_{pq}\subset V_{p'}$, then $d^{(q,p)}_\lambda(A)\geq d^{(q,p')}_\lambda(A)$.
\item if $p$ and $p'$ are not comparable, there is no \emph{a priori} relation between the defect indices.
\eni
On the other hand,
\bei
\item if $q'\leq q$, then $d^{(q,p)}_\lambda(A)\leq d^{(q',p)}_\lambda(A)$.
\item if $q\leq q'$ and $\mathrm{Ran} (A-\lambda I)_{pq'}\subset V_{p}$, then $d^{(q,p)}_\lambda(A)\geq d^{(q',p)}_\lambda(A)$.
\item if $q$ and $q'$ are not comparable, there is no \emph{a priori} relation between the defect indices.
\eni
 \beprop 
  Let $A\in \mathrm{Op}(V_J)$ and  $(q,p)\in  {\sf j}(A)$. Then:
\bei
\item[(i)] $\lambda\in \pi^{(q,p)}(A)$ if and only if $(A-\lambda I)_{pq}$ has a bounded inverse $[(A-\lambda I)_{pq}]^{-1}$ defined on $\mathrm{Ran} (A-\lambda I)_{pq}\subset V_{p}$.
\item[(ii)] $\mathrm{Ran}  (A-\lambda I)_{pq} = \ov{\mathrm{Ran}[ (A-\lambda I)\up V\co]}^{\,p}$ for each $\lambda \in \pi^{(q,p)}(A)$, , where $\ov{\{\cdot\}}^{\,p}$ denotes the closure in $V_p$.

\item[(iii)] if $\lambda \in \pi^{(q,p)}(A)$, then $\mathrm{Ran} (A-\lambda I)_{pq}$ is closed in $V_p$.
\eni
\enprop
\bdim
(i) follows from \eqref{eq:regular}.

(ii) Let $g\in \ov{\mathrm{Ran}[ (A-\lambda I)\up V\co]}^{\,p}$. Thus there exists a sequence $\{f_n\}$ in $V\co$ such that
$g_n :=  (A-\lambda I)f_n \to g\in V_p$.  By \eqref{eq:regular}, we have
$$
 \norm{q}{f_n - f_m} \leq  c_\lambda^{-1} \norm{p}{(A-\lambda I)(f_n - f_m)} = c_\lambda^{-1}\norm{p}{g_n - g_m},
$$
which implies that $\{f_n\}$ is Cauchy in $V_q$, hence convergent in $V_q$. Let $f = \lim_{n}f_n$. Then
$Af_n = g_n + \lambda f_n \to g + \lambda f \in V_p$. Since $A_{pq}$ is continuous, this implies $Af = g + \lambda f $, thus
$(A-\lambda I)f = g \in \mathrm{Ran}  (A-\lambda I)_{pq}$. Hence
$  \ov{\mathrm{Ran}[ (A-\lambda I)\up V\co]}^{\,p} \subseteq \mathrm{Ran}  (A-\lambda I)_{pq}$. The converse inclusion follows from the completeness of  $V_q$.

(iii) follows from (ii).
\edim

The following result follows closely \cite[Props. 3.13 and 3.14]{bello-db-trap}.
 \beprop
 (i) The set   $\pi^J(A)$ of $J$-regular points of $A$ is an open subset of $\CN$.
 \\
 (ii) Let $(q,p)\in  {\sf j}(A)$. Then the defect number $d^{(q,p)}_\lambda(A)$ is constant on each connected component of the open set $\pi^{(q,p)}(A)$.
\enprop
\bdim
(i) Let $\lambda_0 \in  \pi^J(A)$. hence there exist
$(q,p)\in  {\sf j}(A)$, with $q\leq p$ such that $\lambda_0 \in  \pi^{(q,p)}(A)$. Assume that
$| \lambda  - \lambda_0| < c_{\lambda_0}$, where $ c_{\lambda_0}$ is the constant appearing in \eqref{eq:regular}.
Then one shows easily that $\lambda$ satisfies \eqref{eq:regular}, that is, $\lambda\in  \pi^{(q,p)}(A)$.
This in turn implies $\lambda\in  \pi^J(A)$.

As for (ii), this follows from a standard argument.
\edim

\subsection{Resolvents}
\label{subsec-resolv}

Now we turn to resolvents, following the   pattern traditional for \hs s.
Given $(q,p)\in  {\sf j}(A)$, with $q\leq p$, define
 $$
 \rho^{(q,p)}(A): = \{\lambda\in \CN  :  (A-\lambda I)_{pq} \mbox{ is bijective} \}.
$$
Since $(A-\lambda I)_{pq}$ is bijective, it has a continuous inverse $ [(A-\lambda I)_{pq}]^{-1}: V_p \to V_q \in  \B(V_p,V_q)$,  where,
as usual, $ \B(V_p,V_q)$ denotes the set of bounded linear operators from $V_p$ to $V_q$.
Of course,  every $\lambda \in  \rho^{(q,p)}(A)$ is a $J$-regular point of $A$.

 \bedefi \label{def-Jresolvent}
 Let $A\in \mathrm{Op}(V_J)$. Then the \emph{$J$-resolvent set} of $A$, noted $\rho^J(A)$, is the set of complex numbers $\lambda$ for which
  there exists  $(q,p)\in  {\sf j}(A)$, with $q\leq p$, such that     $ (A-\lambda I)_{pq}$ is bijective.
Thus we have:
\be\label{eq-rho}
 \rho^J(A) =\bigcup_{(q,p)\in {\sf j}(A)} \rho^{(q,p)}(A) = \ov{\rho^J(A\ta)},
  \en
where the overbar denotes complex conjugation. The \emph{$J$-spectrum} of $A$ is $\sigma^J(A):=\CN \setminus \rho^J(A)$.
\findefi
For future use, we recall the known facts (see \cite{bello-db-trap} for a proof)
 that the set of invertible elements is open in $ \B(V_p,V_q)$ and that the map
$A \mapsto A^{-1}$ is continuous in $ \B(V_p,V_q)$.

For $\lambda \in\rho^{(q,p)}(A) $, the operator $(A-\lambda I)_{pq}: V_q \to V_p$ is bijective and continuous, hence invertible with the bounded inverse $[(A-\lambda I)_{pq}]^{-1} \in  \B(V_p,V_q)$. Therefore, as discussed in Section \ref{subsec-invert}, the operator
$(A-\lambda I)\in \mathrm{Op}(V_J)$ is invertible, in the sense of Definition \ref{invertible}, with inverse $R_\lambda (A):= (A-\lambda I)^{-1} \in \mathrm{Op}(V_J)$, called the   \emph{$J$-resolvent} of $A$, and  defined by the representative
 $$
R_\lambda (A)_{qp} =   [(A-\lambda I)^{-1}]_{qp} =  [(A-\lambda I)_{pq}]^{-1} : V_p \to V_q\, .
 $$
Clearly, we have
$$
{\sf j}(R_\lambda (A)) = \{ (p,q) \mbox{ such that } (q,p) \in {\sf j} (A)   \mbox{ and } \lambda \in \rho^{(q,p)}(A)\}.
$$

  \medskip
Using these notions, one may prove the  standard results on analytical properties found  in spectral theory, for instance in \cite[Sec.2.2]{schm}. In a first step we fix a suitable pair $(q,p)$.

\beprop\label{prop29}
Let $(q,p) \in  {\sf j}(A) \cap  {\sf j}(B)$, with $q\leq p$. Then
\bei
\item[(i)] $R_\lambda (A)_{qp}- R_\lambda (B)_{qp} = R_\lambda (A)_{qp}\, (B-A)_{pq}\, R_\lambda (B)_{qp},
\;\forall\, \lambda \in  \rho^{(q,p)}(A) \cap  \rho^{(q,p)}(B)$.

\item[(ii)] $R_\lambda (A)_{qp} - R_\mu (A)_{qp}
= (\lambda-\mu) R_\lambda (A)_{qp}\,R_\mu (A)_{qp}, \; \forall\, \lambda,\mu\in \rho^{(q,p)}(A) $.

\item[(iii)] $\rho^{(q,p)}(A)$ is open.

\item[(iv)] The function $\lambda \mapsto  R_\lambda (A)_{qp}\in \B(V_p,V_q)$ is analytic on every connected component of  $\rho^{(q,p)}(A)$ .
\eni
\enprop
For details and proofs, see \cite[Lemma 3.20 and Theor. 3.21]{bello-db-trap}.
 \medskip

For fixed  $q$, if $ \lambda\in  \rho^{(q,p)}(A)$ for some $p\geq q$, this $p$ is unique. Thus we may as well write
\be\label{effq}
f^{(q)}(\lambda):=  R_\lambda (A)_{qp} = [(A-\lambda I)^{-1}]_{qp}\in \B(V_p,V_q)
\en
Hence, according to Proposition \ref{prop29} (iv), the function $f^{(q)}$
 is a single valued function, analytic  in the operator norm of $\B(V_p, V_q)$  on every connected component of the open set $\rho^{(q,p)}(A)$ .

The next step is to      obtain the resolvent series. To that effect, we have to define powers of operators between different spaces (such as the resolvents).
Given $A,B  \in \B(V_p,V_q),\, q\leq p$, define $A_0 := A\up V_q$. Then successive powers may be defined as follows:
\bei
\item $AB f = A_0(Bf), \; \forall\, f\in V_p$, so that $AB : V_p \to V_q$ and  $\norm{p,q}{AB}\leq \norm{q,q}{A_0} \norm{p,q}{B}$;
\item $A^{(2)}: = A_0 A \; \mathrm{and} \; A^{(n)}: = A_0 A^{(n-1)}$.
\eni
 Then, as in \cite[Prop. 3.23]{bello-db-trap}, we get
\beprop  Let $(q,p)\in  {\sf j}(A)$, with $q\leq p$, and $\lambda_0 \in\rho^{(q,p)}(A)$. Then there exists $\delta>0$ such that, for every $\lambda\in\CN$ with
$|\lambda-\lambda_0| < \delta$, $\lambda \in\rho^{(q,p)}(A)$ and
$$
R_\lambda (A)_{qp}= \sum_{n=0}^\infty (\lambda-\lambda_0)^n  [R_{\lambda_0} (A)_{qp}]^{(n+1)},
$$
where the series converges in the operator norm of $\B(V_p, V_q)$.
\enprop

     \medskip
Using the  notations of \pip s, we may rewrite Proposition \ref{prop29} in an intrinsic way.
\betheo \label{theo211}
Let $A, B \in \mathrm{Op}(V_J)$. Then the following statements hold true.
\bei
\item[(i)] Suppose there is a couple $(q,p) \in  {\sf j}(A) \cap  {\sf j}(B)$, with $q\leq p$, such that  $\rho^{(q,p)}(A) \cap  \rho^{(q,p)}(B)
\neq \emptyset$. Then
$$
R_\lambda (A) - R_\lambda (B) = R_\lambda (A)\, (B-A)\, R_\lambda(B),\;\forall\, \lambda \in  \rho^{(q,p)}(A) \cap  \rho^{(q,p)}(B).
$$

\item[(ii)] $R_\lambda (A) - R_{\mu} (A)
= (\lambda-\mu)\,R_\lambda (A) \,R_{\mu} (A), \; \forall\, \lambda,\mu\in \rho^{(q,p)}(A) $.

\item[(iii)] $\rho^J(A)$ is open.

\item[(iv)] The function $\lambda \mapsto  R_\lambda (A)\in \mathrm{Op}(V_J)$ is analytic on every connected component of  the open set  $\rho^J(A)$, for  the inductive topology $\tau_{\rm ind}$ defined by the spaces $\B(V_p, V_q)$.
 \eni
\entheo
 \bdim
(i) The $(q,p)$-representative of the statement is exactly relation (i) of Proposition \ref{prop29}.

(ii) Same argument with relation (ii) of Proposition \ref{prop29}.

(iii)-(iv) The statements  follow from  the corresponding ones of Proposition \ref{prop29}  and \eqref{eq-rho}.
\edim

    \medskip

\berem
Although these results are not more general  at first sight  than the usual, \hs, ones,  Theorem \ref{theo211} shows, once again, how the \pip\ language allows to treat very singular operators as if they were bounded. Examples will be given in Section \ref{sect-ex}.
 \enrem

As a consequence of  (iv),   the resolvent function is clearly analytic with respect to the weak topology of $\mathrm{Op}(V_J)$ defined by the
 seminorms $ X \mapsto |\ip{Xf}{g}|, \; f,g \in V^{\#}.$
 To be precise, we introduce a formal definition.
 \bedefi
The function  $B: z\mapsto B(z)\in \mathrm{Op}(V_J)$ is said to be \emph{weakly analytic} at $z_0\in\CN$ if there exists an operator
$B'(z_0)\in\mathrm{Op}(V_J)$ such that
$$
\lim_{z\to z_0}\left\langle \left.\left({\frac{B(z)-B(z_0)}{z -z_0}- B'(z_0)} \right) f \right| g \right \rangle = 0, \; \forall \, f,g\in V\co.
$$
 \findefi
 But there is more. Under a mild condition of continuity, weak analyticity implies analyticity with respect to the norm of the space $\B(V_p, V_q)$ for some couple $(q,p)\in {\sf j}(A)$. The following proposition applies, in particular, to the resolvent function
 $\lambda \mapsto  R_\lambda (A)\in \mathrm{Op}(V_J)$.

 \beprop\label{prop214}
Let $B: z\mapsto B(z)\in \mathrm{Op}(V_J)$   be  {weakly analytic} in an open set $\A\subset \CN$. Assume there is a couple $(q,p)\in J\times J$ and an open neighbourhood $U(z_0) \subset \A$ such that $(p,q)\in {\sf j}(B(z)), \; \forall\, z\in U(z_0) $. If the function $z\mapsto B(z) f$ is continuous from $\A$ into $V_p$ for every fixed $f\in V\co$, then the function $z\mapsto B(z) _{qp}$ is analytic on $U(z_0)$ with respect to the norm $\norm{p,q}{\cdot}$.
\enprop

First we need a lemma.
\belem
Let the circle $\C:=\{z : |z -z_0| = r\}$ be contained in $U(z_0)$. Then, under the assumptions of Proposition \ref{prop214}, one has
\be\label{eq-bound}
\gamma:=\sup_{z\in \C}\norm{p,q}{B(z) _{qp}}< \infty.
\en
\enlem
\bdim
Since $\C$ is compact, the continuity of the function $ z \mapsto  B(z) f$ from $\A$ into $V_p$ implies there exists a constant
$\gamma_{fg} > 0 $ such that, for all  $z\in\C$,
$$
| \ip{B(z) f}{g} | \leq \gamma_{fg}, \; \forall\,  f\in V\co, \, g\in V_{\ov p}.
$$
Then, by the uniform boundedness theorem, $\sup_{z\in \C}\norm{p}{B(z)f}< \infty$. This in turn implies
the relation \eqref{eq-bound}.
\edim

\textbf{Proof of Proposition \ref{prop214}. }
We investigate the analyticity of the function $U(z_0)\ni z \to B(z) _{qp}$. Putting $\C \subset U(z_0)$ and
$z = z_0 + h$, we have, by the Cauchy integral formula,
\begin{align}
\lefteqn{\left.\left\langle \frac{B(z_0 + h) - B(z_0)}{h}f \right| g\right\rangle - \ip{B'(z_0) f}{g}=}\nn
\\
&\hspace*{1cm}=\frac{1}{2\pi i}\oint_\C \frac{1}{h}\Big(\frac{1}{z-(z_0 +h)} - \frac{1}{z-z_0}\Big) \ip{B(z) f}{g}\ud z
-\frac{1}{2\pi i}\oint_\C \frac{1}{(z-z_0)^2}\ip{B(z) f}{g}\ud z \nn
\\[2mm]
&\hspace*{1cm}=\frac{1}{2\pi i}\oint_\C \left(\frac{1}{h}\Big(\frac{1}{z-(z_0 +h)} - \frac{1}{z-z_0}\Big)
- \frac{1}{(z-z_0)^2} \right) \ip{B(z) f}{g}\ud z. \label{eqiot}
\end{align}
Hence
\begin{align*}
\lefteqn{\left|
\left.\left\langle \frac{B(z_0 + h) - B(z_0)}{h}f \right| g\right\rangle - \ip{B'(z_0) f}{g}
\right| \leq}
\\
&\hspace*{1cm} \frac{1}{2\pi i}\oint_\C
\left|\left(\frac{1}{h}\Big(\frac{1}{z-(z_0 +h)} - \frac{1}{z-z_0}\Big)
- \frac{1}{(z-z_0)^2} \right)\right|  \norm{q,p}{B(z) _{qp}}\norm{p}{f}\norm{\ov q}{g}\ud z.
\end{align*}
In virtue of \eqref{eq-bound}, this implies
$$
\left|
\left.\left\langle \frac{B(z_0 + h) - B(z_0)}{h}f \right| g\right\rangle - \ip{B'(z_0) f}{g}
\right|
\leq \gamma \norm{q}{f}\norm{\ov p}{g}.
$$
Since
$$
\frac{B(z_0 + h) - B(z_0)}{h}f= \frac{B_{qp}(z_0 + h) - B_{qp}(z_0)}{h}f \in V_q,
$$
 \vspace*{0mm}

\noi it follows that $|\ip{B'(z_0) f}{g}| \leq \gamma' \norm{p}{f}\norm{\ov q}{g}.$
This implies that $(p,q)\in{\sf j}(B'(z_0))$.

 From this it follows also that the equality \eqref{eqiot} extends to all $f\in V_p$ and $g\in V_{\ov q}$.
 Thus $z \mapsto B(z)$ is analytic as a map from $U(z_0)$ into $B(V_p,V_q)$. Hence it is analytic with respect to the norm
 $\norm{p,q}{\cdot}$ and $B'(z_0)$ belongs to the same space.
 \hfill $\square$

 \medskip

If $\lambda_0 \in \rho^{(q,p)}(A)$   the function $f^{(q)}(\lambda)$ defined in \eqref{effq}
 is analytic on a disk $D_r = \{\lambda:  | \lambda - \lambda_0 | < r\} \subset \rho^{(q,p)}(A)$ around $\lambda_0$.
In the same way, if $\lambda_0 \in \rho^{(t,s)}(A)$   the function $f^{(t)}(\lambda)$ is analytic on another disk $D_{r '}  \subset \rho^{(t,s)}(A)$ around $\lambda_0$. As discussed in the proof of  Proposition \ref{prop-inverse}, we know that $(q\wedge t,p\wedge s)$
 belongs to ${\sf j}(A)$ and  $A_{p\wedge s, q\wedge t}$ is the restriction of $A_{pq}$ to $V_{q\wedge t}$. In the same way,
 $R_\lambda ^{(q\wedge t,p\wedge s)}(A) $ is the restriction of $R_\lambda ^{(q,p)}(A) $ to $V_{p\wedge s}$.
Hence, the function $f^{(q\wedge t)}(\lambda)$ is the restriction of the function $f^{(q)}(\lambda)$, and also of $f^{(t)}(\lambda)$.
 In other words, these functions are analytic continuations of each other and there is only one resolvent
$R_\lambda(A)= (A-\lambda I)^{-1}\in  \mathrm{Op}(V_J)$ with representatives $R_\lambda ^{(q,p)}(A)$,
each of them analytic on the corresponding open set $\rho^{(q,p)}(A)$.

Next, varying $p$ or $q$ gives rise to (generalized) eigenvalues, which are defined in the obvious way.
 \bedefi Given $A\in \mathrm{Op}(V_J)$, we say that $\lambda$ is a \emph{(generalized) eigenvalue} of $A$ if there is a pair $(q,p)\in  {\sf j}(A)$ such that $A_{pq}- \lambda E_{pq}= (A-\lambda I)_{pq} $ is not injective. Every nonzero vector $f\in \mathrm{Ker}(A-\lambda I)_{pq}$ is called a   \emph{(generalized) eigenvector}.
 If this is true for every $q\in J$, equivalently, for $V_q = V\co$, we say that $\lambda$ is a (global) eigenvalue of $A$.
\findefi

Then one has:
\belem  Let $(q,p)\in  {\sf j}(A), \lambda \in\rho^{(q,p)}(A) $, with $q\leq p$. Then
 \bei
 \item[(i)] $ \lambda\in\sigma^{(q,p')}(A), \forall\, p'\gneqq p $, where $\sigma^{(q,p')}(A):=\CN \setminus \rho^{(q,p')}(A) $ may be called  the relative spectrum of $A$.

 \item[(ii)] if $q \lneqq q' \leq p$ and $(q',p)\in {\sf j}(A)$, then $ \lambda$ is an eigenvalue of $A_{pq'}$ and hence $\lambda\in \sigma^{(q',p)}(A)$.
  \eni
\enlem
\bdim Since $\lambda \in\rho^{(q,p)}(A) $, the operator $(A-\lambda I)_{pq}$ is bijective from $V_q$ onto $V_p$. Thus it cannot be bijective onto
$V_{p'}$, which means that $\lambda\not \in\rho^{(q,p')}(A) $, i.e., $ \lambda\in\sigma^{(q,p')}(A)$.

(ii) Again $(A-\lambda I)_{pq}$ is bijective from $V_q$ onto $V_p$. Take $f'\in V_{q'}\setminus V_q$. Since $(A-\lambda I)_{pq}$ is surjective, there exists a unique vector $f \in V_{q}$ such that $(A-\lambda I) f' = (A-\lambda I) f$, hence $g=f'-f \in V_{q'}$ is nonzero and satisfies
$(A-\lambda I) g=0$, thus $ \lambda$ is an eigenvalue of $A_{pq'}$ and hence $\lambda\in \sigma^{(q',p)}(A)$.
\edim

\subsection{RHS generated by symmetric operators: A counterexample}
\label{subsec-ex}

Given a self-adjoint operator  $A$ in the \hs\ $\H$, the scale built on the powers of $A$ is constructed in the standard way.
For $n\in \NN$, define $\H_n = D(A^n)$ with the graph norm $\norm{n}{\cdot} = \norm{}{(I+A^{2n})^{1/2}}$ and
$\H_{\ov n}:= \H_{-n}$ as the completion of $\H$ with respect to the norm $\norm{\ov n}{\cdot} = \norm{}{(I+A^{2n})^{-1/2}}$.
Then $V_J = \{\H_n, n\in\ZN\}$ is the familiar scale \cite[Sec.5.2.1]{pip-book}:
\begin{equation}
\label{eq:scaleHn}
D^\infty(A):=\bigcap_{n} \H_n \subset  \ldots \subset   \H_2  \subset   \H_1  \subset  \H_0 \subset
  \H_{\ov 1}   \subset \H_{\ov 2} \ldots \subset  \bigcup_{n} \H_{n}.
\end{equation}
In this scale,
$A$ maps $\H_n$ into $\H_{n-1}$ continuously, for every $n\in\ZN$. Denote by $\rho_\H(A)$ the usual resolvent of $A$. Let $S\in \mathrm{Op}(V_J)$ denote the operator defined by $A\up D^\infty(A)$.
Then it is shown in \cite[Prop. 4.1]{bello-db-trap} that $\rho^J(S) = \rho_\H(A)$. This result, however, fails in  a more complicated case.

Let indeed  $S$ be a closed symmetric operator with several self-adjoint extensions $S_\alpha, \alpha\in I$.
For any self-adjoint extension $S_\alpha$ of $S$, $D(S^n)\subset D(S_\alpha^n)$ and
$D(S^\infty)\subset D(S_\alpha^\infty)$.

Put $\H_{\alpha,n}= D(S_\alpha^n)$, with the graph norm, and $V_ {J_0} = \{\H_{\alpha,n}, \alpha\in I, n\in\NN\}$.
Then $S: \H_{\alpha,n}\to \H_{\beta,m}$ continuously if and only if  $\alpha=\beta$ and $m\leq n-1$.
Next, for each $\alpha\in I, \rho^{(\H_{\alpha,n},\H_{\alpha,n-1})} = \rho_\H(S_\alpha)$ and
$\rho^{J_0}(S)= \bigcup_{\alpha\in I}\rho_\H(S_\alpha)$.

  A standard example, given in \cite{bello-db-trap}, is that of the first order differential operator $S:=-i {\ud}/{\!\ud x}$  on a segment of the real line.
  We sketch it here. Define the operator $S$ on the domain
  $$
  D(S) = \{f\in L^2(0,1) : f' \in L^2(0,1), f(0) = f(1) = 0\}.
  $$
Its adjoint is $S^*:=-i {\ud}/{\!\ud x}$ on the domain
  $
  D(S^*) = \{f\in L^2(0,1) : f' \in L^2(0,1)\}.
  $
 Next define the operator $S_\alpha  =-i {\ud}/{\!\ud x}$ on the domain
   $$
  D(S_\alpha) = \{f\in L^2(0,1) : f' \in L^2(0,1),   f(1) = \alpha f(0), \alpha\in \CN, |\alpha| = 1\}.
$$
Clearly  one has $S\subset  S_\alpha \subset S^*$,  $S$ is closed and symmetric,
  each $S_\alpha$   is self-adjoint, thus a self-adjoint extension of $S$, but $S_\alpha, S_\beta$ are  not comparable for $\alpha \neq \beta$.
  Then it is shown in \cite{bello-db-trap} that the resolvents $(S_\alpha - \lambda I)^{-1}$ and $(S_\beta- \lambda I)^{-1}$ are \emph{not} analytic continuation of each other if $\alpha \neq \beta$.

Actually, this result does not contradict that of Section \ref{subsec-resolv}, because $V_ {J_0}$ is \emph{not} a LHS, indeed
$J_0$ is not a lattice, since  $\H_{\alpha,n}\cap \H_{\beta,m} \not \in V_ {J_0}$ ! And the lattice property is the key to the uniqueness of  inverse operators, as we have seen in Section \ref{subsec-invert}.
In fact, $V_ {J_0}$  is  a collection of scales of \hs s, $V_{S_\alpha}, \alpha\in I$, mutually incompatible.
To get a genuine LHS, one has to consider the lattice $J$ generated by $J_0$, but this is not very natural \ldots.
For instance,
$$
\H_{\alpha,n}\cap \H_{\beta,m}=D(S_\alpha^n) \cap D(S_\beta^m) \supset D(S ^n) \cap D(S^m) = D(S^{\max(n,m)}).
$$
 Thus it is not surprising that differential operators on an interval yield pathologies (multivalued analytic functions).

\section{Spectral analysis of Hilbert space operators}
\label{sec-hilbert}

In this section, we shall discuss the spectral analysis of self-adjoint operators in \hs s, in the light of \pip s. We consider again the simplest case, namely, a Hilbert scale.

\subsection{Generalized eigenvalues and generalized eigenvectors}

Let $A>I$ be a self-adjoint operator in the \hs\ $\H$ and $V_J = \{\H_n, n\in\ZN\}$ the usual scale on powers of $A$, thus $\H_0= \H$.
Given $X\in \mathrm{Op}(V_J),$ there exists $m\in\NN$ such that $(m,-m)\in {\sf j}(X)$, with $\H_m \subset \H \subset\H_{\ov{m}}$, where
$\H_{\ov{m}}:=\H_{-m}$.
Define
\begin{align}
D(X_{0,m}) &= \{ f\in \H_m : Xf\in \H\} \nn \\
X_{0,m}f &= Xf,\, f \in D(X_{0,m}). \label{defXOM}
\end{align}
Then $X_{0,m}$ is a restriction of $X$, with $D(X_{0,m}) \subset \H_m$, but its domain need not be dense in $\H$, it could even be reduced to $\{0\}$.

In order to go further, we have to resort to a version of the KLMN Theorem \cite[Theorems 3.3.25 and 3.3.27]{pip-book}.

\beprop  \label{prop31}
Let $X = X\ta \in \mathrm{Op}(V_J)$ be a symmetric operator. Assume that $(m,n)\in {\sf j}(X)$, with
 $\H_{m} \subseteq\H_0\subseteq \H_{n}$, and there is a $\lambda$ such that $(X -\lambda I)_{nm} $ is a bijection, hence it is boundedly invertible.
Then there exists a unique restriction of $X_{nm}$
 to a self-adjoint operator  $X_{0}$ in the Hilbert space $\H_0$. The number $\lambda$  does not belong to the (Hilbertian) spectrum of $X_{0}$.
  The domain of  $X_{0}$  is obtained by eliminating from $\H_{m} $ exactly the vectors $f$ that are mapped by $X_{nm}$ beyond $\H_0$.
\enprop
\bdim
Define $D(X_{0})  = \{ f\in \H_m : Xf\in \H_0\}$ and $X_{0}:=X_{nm} \up D(X_{0})$  Define
$$
R_{mn}:=  (X_{nm} -\lambda  E_{nm})^{-1}  : \H_n \to \H_{m}
$$
  as the inverse of the invertible representative $(X -\lambda I)_{nm}$. Then
  $R_{00} = E_{0m}\,R_{mn}\,E_{n0}$
 is a restriction of $R_{mn}$. By \cite[Lemma 3.3.26]{pip-book}, $R_{00}$ has a self-adjoint inverse $(R_{00})^{-1} = (X_{0}-\lambda I)$, which  is a restriction of
$X_{nm} -\lambda  E_{nm}$.  Thus $X_{nm}$ has a self-adjoint, densely defined, restriction to $\H_0$.
Since $R_{mn}$ is bounded, so is $R_{00}= (X_{0}-\lambda I)^{-1}$, thus $\lambda$  does not belong to the spectrum of $X_{0}$.
\edim

This result applies, in particular, to the  case   $(m,\ov{m})\in {\sf j}(X), \, m\in \NN$,   described at the beginning of the section, since $m \leq 0 \leq \ov{m}$.
Then $X$ has an invertible representative $X_{\ov{m} m}$.

{Actually one can go further, in the case of an arbitrary LHS, using the generalized KLMN Theorem \cite[Theorem 3.3.28]{pip-book}. Thus we consider a symmetric operator  $X = X\ta \in \mathrm{Op}(V_J)$ in an arbitrary LHS $V_J = \{\H_{n}, n\in J\}$ and we assume   there exists a $ \lambda \in \RN$ such that $X  - \lambda I $ has
an invertible representative $X_{nm} - \lambda E_{nm} : \H_{m} \to \H_{n}$,  where $\H_{m} \subseteq \H_{n}$, but neither of these need be comparable to $\H_0$.
Before stating a proposition and sketching its proof, it is worth clarifying the position of the various spaces involved.

On one hand, the key step in the proof of \cite[Theorem 3.3.28]{pip-book} is the relation
\be\label{triplet-mm}
\H_{m\wedge \ov{m}} \subset \H_0 \subset \H_{m\vee \ov{m}} .
\en
On the other hand, since $X$ is symmetric and $(X - \lambda I)_{nm} :  \H_{m} \to \H_{n}$ is one-to-one and continuous, so is $(X - \lambda I)_{{\ov m},{\ov n}} : \H_{\ov n} \to \H_{\ov m}$, and therefore, by Lemma 3.3.29 of \cite{pip-book}, also
\mbox{$(X - \lambda I)_{{\ov m}\wedge n,m \wedge{\ov n}} : \H_{m \wedge{\ov n}} \to \H_{{\ov m}\wedge n}$}
and $(X - \lambda I)_{{\ov m}\vee n,m \vee{\ov n}} : \H_{m \vee{\ov n}} \to \H_{{\ov m}\vee n}$.
However, since  $m\leq n$, we have also ${\ov n}\leq {\ov m}$ and $m \wedge{\ov n} \leq {\ov m}\wedge n$ and
$m\vee \ov{m} \leq {\ov m}\vee n$.
Therefore, by \cite[Prop.2.5.1]{pip-book}, we can complete  \eqref{triplet-mm}:
\be\label{quintet}
\H_{m\wedge \ov{n}} \subset \H_{m\wedge \ov{m}} \subset  \H_0\subset \H_{m\vee \ov{m}}
\subset \H_{{\ov m}\vee n} .
\en
However, neither $\H_{{\ov m}\wedge n}$, nor $\H_{m \vee{\ov n}}$ need be comparable to $\H_0$.
Thus we take a predecessor, resp. a successor, and consider the map
$(X - \lambda I)_{{\ov m}\vee n,m \wedge{\ov n}} : \H_{m \wedge{\ov n}} \to \H_{{\ov m}\vee n}$.

Now, in addition to \eqref{quintet}, we have two more chains, which do not contain $\H_0$:
\begin{align}
&\H_{m\wedge \ov{n}} \subset \H_{m\wedge \ov{m}} \subset \H_{n \wedge{\ov  m} }\subset \H_{\ov{m}}\subset  \H_{m\vee \ov{m}} \subset\H_{{\ov m}\vee n}\, , \label{eq44}
\\
&\H_{m\wedge \ov{n}} \subset \H_{m\wedge \ov{m}} \subset \H_{{m}}\subset  \H_{m \vee{\ov n}}\subset  \H_{m\vee \ov{m}} \subset\H_{{\ov m}\vee n}\, .
 \label{eq45}
 \end{align}
This is useful when  $V_J$ is a chain, since then $m\wedge \ov{n} = \min(m,\ov{n}), {\ov m}\vee n = \max({\ov m}\vee n)$, and so on. In that situation, we have exactly four cases:
\bee
\item[(i)] If $m \leq \ov{n}$, \eqref{eq44}  yields one of the following
\bei
\item [(ia)]   $m \leq n \leq 0 \leq {\ov n} \leq {\ov m}$ and $X - \lambda I : \H_{m} \to  \H_{\ov m}$,  if $n \leq \ov{n}$,

\item [(ib)] $m \leq {\ov n} \leq 0 \leq n \leq {\ov m}$ and  $X - \lambda I : \H_{m} \to  \H_{\ov m}$, if $n \geq \ov{n}$.
\eni

\item [(ii)] If $m \geq \ov{n}$, \eqref{eq45} yields  yields one of the following
\bei
\item [(iia)] ${\ov n} \leq m \leq 0 \leq  {\ov m} \leq n$ and $X - \lambda I : \H_{\ov n} \to  \H_{n}$, if $m \leq \ov{m}$,

\item [(iib)]  ${\ov n} \leq {\ov m} \leq 0 \leq m  \leq n$ and $X - \lambda I : \H_{\ov n} \to  \H_{n}$, if $m \geq \ov{m}$.
\eni
\ene
Of course, if $m = \ov{n}$, both (ib) and (iia) yield simply $m   \leq 0   \leq {\ov m}$.
In all cases, the triplet \eqref{triplet-mm} reduces either to $\;\H_{m} \subset \H_0 \subset \H_{\ov{m}}\;$  or to $\;\H_{\ov{m}} \subset \H_0 \subset \H_{m}$.

Now we can state a proposition.
\beprop  \label{prop32}
Let $X = X\ta \in \mathrm{Op}(V_J)$ be a symmetric operator in an arbitrary LHS $V_J = \{\H_{n}, n\in J\}$.
Assume   there exists a $ \lambda \in \RN$ such that $X  - \lambda I $ has
an invertible representative $X_{nm} - \lambda E_{nm} : \H_{m} \to \H_{n}$,  where $\H_{m} \subseteq \H_{n}$.
Then  $X_{nm}$ determines a unique, densely defined,
 self-adjoint operator  $X_{0}$ in the Hilbert space $\H =\H_{0}$. The number $\lambda$  does not belong to the spectrum of  $X_{0}$.
\enprop}
\noi{\bf Idea of the proof } \cite[Theorem 3.3.28]{pip-book} :  Let again
$$
R_{mn}:=  (X_{nm} -\lambda  E_{nm})^{-1}  : \H_{n} \to \H_{m}.
$$
This defines a symmetric operator $R=R\ta\in \mathrm{Op}(V_J)$.
The key fact in the proof is the relation \eqref{triplet-mm}.
Then  the three representatives $R_{m\wedge \ov{m},m\wedge \ov{m}}, R_{00}, R_{m\vee \ov{m},m\vee \ov{m}}$ exist,   are injective and have dense range.
Next $R_{00}$ is  self-adjoint in $\H_0$. Since it is injective and has dense range,  its inverse
$(R_{00})^{-1} = X_{0} - \lambda I$ is also self-adjoint and densely defined.
The rest is as in Proposition \ref{prop31}.
\qed

Note that the domain of $X_{0}$ is $D(X_{0}) =R_{00} \H_0$, but we can't say more at this level of generality. As indicated above, we can consider the map
 $(X - \lambda I)_{{\ov m}\vee n,m \wedge{\ov n}} : \H_{m \wedge{\ov n}} \to \H_{{\ov m}\vee n}$ and its restriction to the domain
 $D_{0}: =  \{ f\in \H_{m\wedge\ov{n}} : Xf\in \H_0\}$. However we don't know if this domain is dense in $\H_0$.

In the case of a chain, things get simpler.
\becor\label{cor33}
Let the LHS $V_J = \{\H_{n}, n\in J\}$ of Proposition \ref{prop32} be a chain of \hs s. Then,  the domain of the self-adjoint operator $X_{0}$ may be described explicitly and is obtained by restriction, as in Proposition \ref{prop31}.
\encor
\bdim
In the four cases above, the dual pair
$ \H_{m \wedge{\ov n}},\H_{{\ov m}\vee n}$ reduces to $ \H_{m},\H_{{\ov m}}$ or $ \H_{{\ov n}},\H_{n}$ and $X  - \lambda I $ maps the small space bijectively on the large one. Next, in cases (ib) and (iia), one has $ \H_{\ov n}\subset \H_0 \subset\H_{\ov m}$ and $X_{{\ov m},{\ov n}}$ is invertible,
so that Proposition \ref{prop31} applies. In particular the domain of $X_{0}$ is $D(X_{0})=  \{ f\in \H_{{\ov n}} : Xf\in \H_0\}$.

In case (ia), consider the operator $X_{{\ov m},0}$, restriction to $\H_0$ of  $X_{{\ov m},{\ov n}}$. The domain of $X_{0}$ is obtained by restriction of
$X_{{\ov m},0}:  \H_0 \to  \H_{\ov m}$, namely,
$D(X_{0})=  \{ f\in \H_{0} : Xf\in \H_0\}$. Thus one has $\H_{m}\subset D(X_{0}) \subset \H_0 $, which confirms that $D(X_{0})$ is dense in $ \H_0 $.
The case (iib) is similar, passing to the dual spaces, except for the last statement.
\edim

 \medskip

When performing the spectral analysis of a self-adjoint operator in a \hs, the standard tool is the RHS formulation due to Maurin-Gel'fand
\cite[Chap. 1, \S 4]{gelfand4} and also generalized by Roberts \cite{roberts} and one of us \cite{jpa-Dirac1}. In \cite[Theor.4.4]{bello-db-trap}, a slightly more general version was given, which runs as follows.
\betheo \label{theo34}\cite{bello-db-trap}
Let $V_J$  be the scale built on the powers of the self-adjoint operator $A\geq I$,whose inverse $A^{-1}$ is Hilbert-Schmidt. Assume that
$(m,\ov{m})\in {\sf j}(X)$ for some $m\in\NN$ and that $X_{0,m}$, defined in \eqref{defXOM}, is densely defined and essentially self-adjoint.
Then  $X$ has a complete set of generalized eigenvectors belonging to  $\H_{\ov{m}}$.
\entheo

\medskip

\noi \textbf{Remark :}    If  $A^{-1}$ is Hilbert-Schmidt, then $A$ must have a purely point spectrum with finite multiplicity and $V\co = D^\infty(A)  $ is a nuclear Fr\'echet space.

It remains to combine Theorem \ref{theo34} with the previous Proposition  \ref{prop32} to get a genuine generalization of the original Maurin-Gel'fand theorem.
\betheo \label{theo35}
Let $V_J$  be the scale built on the powers of the self-adjoint operator $A\geq I$,whose inverse $A^{-1}$ is Hilbert-Schmidt.
Let $X = X\ta \in \mathrm{Op}(V_J)$ be a symmetric operator. Assume that $(m,n)\in {\sf j}(X)$, with  $\H_m \subset \H_o\subset \H_n$,
 and  there is a $\lambda\in \RN$ such   that $X -\lambda I $ has an invertible representative  $(X -\lambda I)_{nm} $.
 Then, if either  $m \leq {\ov n}  \leq n $  or  ${\ov n} \leq m   \leq  {\ov m}$,
 $X_{nm}$ has a  unique restriction to a self-adjoint operator  $X_{0}$ in $\H$.
In addition, possesses a complete set of generalized eigenvectors belonging to  $\H_{\ov{m}}$, if $m \leq \ov{n}$, or to $\H_{n}$, if $m \geq \ov{n}$.
\entheo
\bdim
The assumption $m \leq {\ov n}  \leq n $  or  ${\ov n} \leq m   \leq  {\ov m}$ means that we are in cases (ib), resp. (iia), among
 the four cases described above.

Let us proceed with   case (ib).   We have $\;\H_{m} \subset \H_0 \subset \H_{\ov m}$ and $X  : \H_{m} \to  \H_{\ov m}$. The domain of $X_{0}$ is
$D(X_{0})=  \{ f\in \H_{m} : Xf\in \H_0\}$, so that we can apply Theorem \ref{theo34}.

Let $\{E(\lambda)\}$ be the spectral family of the self-adjoint operator $X_{0}$ defined in Proposition \ref{prop32} and Corollary \ref{cor33},
 Given  a unit vector $h \in\H$, put $\sigma(\lambda) = \ip{E(\lambda)h}{h}$. Then, in virtue of \cite[Ch. IV, Sect.4.3, Theor.1]{gelfand3}, $\sigma$  defines a (Lebesgue-Stieltjes) measure on $\RN$, such that the following derivative exists almost everywhere:
$$
\chi_\lambda:= \frac{\ud E(\lambda)h}{\ud\sigma(\lambda)} \; \mathrm{a.e.}.
$$
Then $\chi_\lambda$ is a continuous conjugate linear functional on $V\co = D^\infty(A) $, acting as
$$
 \chi_\lambda(f) \equiv \ip{\chi_\lambda}{f}= \frac{\ud \ip{E(\lambda)h}{f}}{\ud\sigma(\lambda)} \; \mathrm{a.e.},  \; f\in D^\infty(A).
$$
This functional is  a generalized eigenvector of   $X$, corresponding to the (generalized) eigenvalue $\lambda$. These generalized eigenvectors form a complete system in the sense of Maurin-Gel'fand, that is, one has, for every $f,g\in V\co$,
\begin{align*}
f & =  \int  \ov{ \ip{\chi_\lambda}{f} }\; \chi_\lambda \ud\sigma(\lambda)\, ,
\\
\ip{f}{g} &= \int  \ov{ \ip{\chi_\lambda}{f}}\; \ip{\chi_\lambda}{g} \ud\sigma(\lambda)\, .
\end{align*}
As for the generalized eigenvectors, the argument of \cite[Theor.4.4]{bello-db-trap} show that they belong to  $\H_{\ov m}$ and form a complete set.

The case (iia) is entirely similar, with $\;\H_{\ov n} \subset \H_0 \subset \H_{n}$, so that
the operator $X$ has now a complete set of generalized eigenvectors belonging to $ \H_{n}$.
\edim
\medskip

As for the two other cases, we note
\bei
\item[(ia)] Here $D(X_{0})=  \{ f\in \H_{0} : Xf\in \H_0\}$ and $\H_{m}\subset D(X_{0})$, whereas $X : \H_{m} \to  \H_{\ov m}$; thus the domain
$D(X_{0m})  = \{ f\in \H_m : Xf\in \H_0\}$ is contained in $D(X_{0})$, and we don't know whether $X_{0}\up D(X_{0m}) $ is essentially self-adjoint, so that Theorem \ref{theo34} may not apply.
\item[(iib)] Same situation as (ia): $X  : \H_{\ov n} \to  \H_{n}$ and the domain $D(X_{0{\ov n}})  = \{ f\in \H_{\ov n} : Xf\in \H_0\}$ is contained in $D(X_{0})$.
\eni

\subsection{Tight riggings}

   Let us go back to the scale built on the powers of the self-adjoint $A$.
   Let $m\geq 0 \geq n$ in Prop. \ref{prop32}, i.e., there exists  $ \lambda \in \RN$ such that
  $X_{nm} - \lambda E_{nm} : \H_{m} \to \H_{n}$ is bijective, hence $[X_{nm} - \lambda E_{nm}]^{-1}$ is bounded. By  Prop. \ref{prop32}, $X_{nm}$ has a unique restriction $X_{0}$ which is a  self-adjoint operator in $\H_0$.
   Its domain is $D(X_{0})  = \{ f\in \H_m : Xf\in \H\}$ and $\lambda\in\rho(X_{0})$.
   Of course, if $n=0, D(X_{0}) = \H_m$.

   Let now $\K \subset D(X_{0})$ be defined as $\K := \{f \in D(X_{0}): X_{0}f \in \H_m\}$. Put on $\K$ the graph norm of $X_{0}$ in $\H_m$:
   $$
   \norm{\K}{g}^2:=    \norm{m}{g}^2 + \norm{m}{X_{0}g}^2.
   $$
Thus, assuming that $K$ is dense in $\H$, we have
$$
\K  \subset   D(X_{0})  \subseteq \H_m \subset \H \subset \H_{\ov m} \subset \K \ta
$$
  and $ X_{0}$ maps $\K$ continuously into $\H_m$. In other words, the pair $(\K, \H_m)$ is admissible with respect to $ X_{0}$, or is a rigging for $ X_{0}$, in the sense of Babbitt \cite{babbitt} or Berezanskii \cite[Chap. V, \S 2]{berez}.

   Let now $X_{0}^\dag : \H_{\ov m} \to \K \ta$ be the adjoint of $X_{0}$, defined by
   $$
   \ip{g}{X_{0}^\dag\Psi} = \ip{X_{0}g}{\Psi}, \; \mathrm{for}\; g\in \K, \Psi\in \H_{\ov m} \, .
   $$
   Then $\Psi\in  \H_{\ov m}$ is a \emph{generalized eigenvector} for $X_{0}$, with \emph{generalized eigenvalue}
   $\lambda$,  if $X_{0}^\dag\Psi = \lambda\Psi$.

Define $\sigma_\mathrm{ext}(X_{0}) :=  \sigma_{\K, \H_{m}}(X_{0}) \supset \sigma(X_{0})$, the \emph{extended spectrum}  of  $X_{0}$, as the closure of the set of all generalized eigenvalues of $X_{0}$ for the rigging  $(\K, \H_m)$.
Comparing with Definition  \ref{def-Jresolvent}, we get
$$
\sigma(X_{0}) \subseteq \sigma_\mathrm{ext}(X_{0}) \subseteq \sigma^J(X).
$$
Then one says that the rigging is \emph{tight} if the two spectra $\sigma$  and $\sigma_\mathrm{ext}$ coincide, that is, the rigging does not bring in new eigenvalues.
To give a classical example, take $X_0 = - i \frac{\ud}{\ud x}$ in $L^2(\RN)$. In the Schwartz RHS $\cS \subset L^2\subset   \cS\ta$, one gets a tight rigging, since
$\sigma(X_{0}) =\sigma_\mathrm{ext}(X_{0}) = \RN$ with generalized eigenvectors $\Psi_\lambda (x)= e^{i \lambda x} \in  \cS\ta$.
But if one takes the RHS $\D \subset L^2\subset   \D\ta$, where $\D$ now denotes the Schwartz space of $C^\infty$ functions of compact support,
one obtains $\sigma_\mathrm{ext}(X_{0}) = \CN$, with the same eigenvectors, which no not belong to $ \cS\ta $ if $\mathrm{Re}\, \lambda \neq 0$.

 In this context, Babbitt \cite[Lemma 2.2] {babbitt} shows that the rigging $(\K, \H_m)$  is tight if and only if
 $(X_{0} - \lambda I)\K$ is dense in $\H_m$ for every $\lambda\in\rho(X_{0})$.

 In the general case  of an arbitrary LHS and  $\H_{m} \subseteq \H_{n}$ in Prop. \ref{prop32}, one gets
 $$
\K \subset   D(X_{0})  \subseteq \H_{m\wedge \ov{n}} \subset \H_0 \subset \H_{n\vee \ov{m}} \subset \K \ta,
$$
where $\K := \{f \in D(X_{0}): X_{0}f \in \H_{m\wedge \ov{m}}\}$, with the graph norm from
$\H_{m\wedge \ov{m}}$. One then proceeds as before.

 \medskip

Now we pose the following question: Given $X=X\ta \in \mathrm{Op}(V_J)$, is it possible to construct a generalized resolution of the identity?

{As we have seen in Theorem \ref{theo35}, to certain symmetric operators $X$ of $\mathrm{Op}(V_J)$ there corresponds a complete family of generalized eigenvectors $\{\chi_\lambda\}\subset V$  in the sense that there exists a positive Borel measure $\sigma$ on the real line such that
\begin{itemize}
\item[(i)] $\dis\int_{\mb R} |\ip{\chi_\lambda}{f}|^2 \ud\sigma(\lambda) <\infty, \quad \forall\, f \in V\co$,
\item[(ii)] $f=\dis\int_{\mb R} \overline{\ip{\chi_\lambda}{f}}\chi_\lambda \ud\sigma(\lambda) , \quad \forall \,f \in V\co$,
\item[(iii)] $\ip{f}{g}=\dis\int_{\mb R} \overline{\ip{\chi_\lambda}{f}}\ip{\chi_\lambda}{g} \ud\sigma(\lambda) , \quad \forall \,f,g \in V\co$,
\item[(iv)] $\ip{\chi_\lambda}{Xf}= \lambda \ip{\chi_\lambda}{f}, \quad \forall\, f \in V\co$.
\end{itemize}

Assume now that $X$ is a symmetric operator of $\mathrm{Op}(V_J)$ possessing a complete family $\{\chi_\lambda\}$ of eigenvectors (i.e., (i)-(iv) hold) corresponding to real generalized eigenvalues $\lambda$'s.
If $X$ maps $V^{\#}$ into itself, then by (iii) and (iv), we get, for $f,g \in V\co$,
\begin{align}\label{eq_spec}
\ip{Xf}{g}&=\int_{\mb R} \overline{\ip{\chi_\lambda}{Xf}}\ip{\chi_\lambda}{g} \ud\sigma(\lambda) \\
&=\int_{\mb R} \lambda\, \overline{\ip{\chi_\lambda}{f}}\ip{\chi_\lambda}{g} \ud\sigma(\lambda). \nonumber
\end{align}
This fact suggests the possibility of defining, as in the discrete case, an operator $B(\mu)$, $\mu \in {\mb R}$, by
$$
B(\mu)f= \int_{\alpha\leq \mu}\overline{\ip{\chi_\alpha}{f}}\chi_\alpha \ud\sigma(\alpha), \quad f \in V\co.
$$
Then each $B(\mu)$ is a symmetric element of $\mathrm{Op}(V_J)$. Indeed,  it is a bounded symmetric operator in  $V_0$, since it satisfies the relation
$\ip{B(\mu)f}{f}\leq \norm{}{f}^2, \;\forall\, f\in V\co$.
Moreover, it is not difficult to check that the family $\{B(\mu)\}$  has the   same properties of an ordinary spectral family, with the possible exception of idempotence. By the definition itself it follows that
\begin{align*}
\overline{\ip{\chi_\lambda}{f}}\ip{\chi_\lambda}{g} \ud\sigma(\lambda)&=\ud\, \int_{\alpha\leq \mu}\overline{\ip{\chi_\alpha}{f}}\chi_\alpha \ud\sigma(\alpha)\\
&=\ud\ip{B(\lambda)f}{g}, \quad  \forall \,f, g \in V\co.
\end{align*}
This allows to rewrite \eqref{eq_spec} in a more familiar form:
$$
\ip{Xf}{g} = \int_{\mb R} \lambda \ud\ip{B(\lambda)f}{g}, \quad  \forall\, f, g \in V\co.
$$
The assumption $XV^{\#} \subseteq V^{\#}$ can be weakened in an obvious way. Indeed, if $X:V\co \to V_r$, $r\leq 0$, generalized eigenvectors, if they exist, live necessarily in $V_{\ov{r}}$ and a set of vectors $\{\chi_\lambda\}$ will be called complete if
\begin{itemize}
\item[(i')] $\dis\int_{\mb R} |\ip{\chi_\lambda}{f}|^2 \ud\sigma(\lambda) <\infty, \quad \forall \,f \in V_r$,
\item[(ii)] $f=\dis\int_{\mb R} \overline{\ip{\chi_\lambda}{f}}\chi_\lambda \ud\sigma(\lambda) , \quad \forall \,f \in V_r$,
\item[(iii)] $\ip{f}{g}=\dis\int_{\mb R} \overline{\ip{\chi_\lambda}{f}}\ip{\chi_\lambda}{g} \ud\sigma(\lambda) , \quad \forall \,f,g \in V_r$.
\end{itemize}

In conclusion, we have
\begin{prop} Let $X=X^\times \in \mathrm{Op}(V_J)$. Assume that $X:V\co \to V_r$, $r\leq 0$ and that $X$ has a complete family $\{\chi_\lambda\}\subset
V_{\ov{r}}$ corresponding to real generalized eigenvalues $\lambda$'s.
Then there exists a generalized spectral family $\{B(\mu)\}$ such that
$$
\ip{Xf}{g} = \int_{\mb R} \lambda \ud\ip{B(\lambda)f}{g}, \quad  \forall f \in V^{\#}, g \in V_{\ov{r}}.
$$
\end{prop}}
\noi We may remark that the family $\{B(\mu)\}$ is a generalized resolution of the identity in the sense of Na\u{\i}mark \cite[Appendix]{akhiezer} or
 \cite[Appendix]{riesz}.

\section{Examples}
\label{sect-ex}

We give here some simple examples of singular symmetric operators in a LHS and discuss their spectral properties.

\beex {\bf Singular interactions in quantum mechanics}\label{ex41}
\\[2mm]
A typical Hamiltonian in (nonrelativistic) quantum mechanics takes the form $H= H_0 + V$, where the potential is the operator of multiplication $M_V$ by the function $V$. If one chooses for potential a $\delta$ function, one gets a singular (zero range or point) interaction and the problem is to give a precise meaning to the symbolic expression : $H ``\!=\!" H_0 + \delta$.

Such singular interactions are   discussed in \cite[Sec.7.1.3]{pip-book}, namely,
 the description  of quantum mechanical systems with local, many-center Hamiltonians, based on the original paper of Grossmann \emph{et al.} \cite{ghm}.
We give here a simplified version of that example.
Let $V = L^1_{\rm loc}(\RN^n, \ud p)$ and let $T$ be multiplication operator by the positive unbounded function  $t(p)$, so that $\lambda=-1$ is
 a real point in the resolvent set of $T$ (typically $t(p) = p^2$, that is,  $T$  is the free Hamiltonian). Then consider the scale $V_T:= \{\H_{r}, r\in\ZN\}$ built on the powers of  $(T+I)^{1/2}$, where
\be\label{eq:scale-T}
\H_{r}(\RN^\nu) := \{\phi \in  L^1_{\rm loc}(\RN^\nu): \|\phi\|^2_{r}:= \int_{\RN^\nu} \big(t(p)+1 \big)^r | \phi (p)|^2\ud p < \infty\}.
\en
For $t(p) = p^2$, we get again the Sobolev spaces (or their Fourier transforms), so the  scale \eqref{eq:scale-T} is again  not nuclear. Thus Theorem \ref{theo35} does not apply.

In particular, we will use the central part of the scale \eqref{eq:scale-T}, namely,
\be\label{eq:finitescale-T}
\H_{2} \subset \H_{1} \subset\H_{0} \subset\H_{\ov 1} \subset\H_{\ov 2}\ ,
\en
where as usual $\H_{\ov r} = \H_{-r}$.
The free resolvent is the operator $ R_ \lambda(T)= (T-\lambda I)^{-1}$, that is, the operator of multiplication by $(t(p)-\lambda)^{-1}$, where $\lambda$ belongs to the resolvent set $\rho(T)=\CN \setminus [0,\infty)$. Then,

\noi(i) Viewed as an operator in the central \hs\ $\H_{0}$, $ R_ \lambda(T)$ satisfies the identities
\begin{align}
 R_ \lambda(T)- R_ {\mu}(T) &= (\lambda-\mu) R_ \lambda(T) R_ {\mu}(T),  \label{eq:ident1}\\[1mm]
\frac{\ud  R_ \lambda(T)}{\ud\lambda} &=  R_ \lambda(T)^2. \label{eq:ident2}
\end{align}

\noi(ii)  $ R_ \lambda(T): \H_{r}\to \H_{r+2}$ is bounded with bounded inverse, 
and similarly for   $ R_ \lambda(T)^{1/2}: \H_{r}\to \H_{r+1}$.
Therefore, $\ip{f}{ R_ \lambda(T)g}$ is well-defined for $f,g\in \H_{\ov 1}$, and
$\ip{f }{ R_ \lambda(T)^2g }$ is well-defined for $f,g \in \H_{\ov 2}$.
Clearly, these statements are  in accordance with Proposition \ref{prop29} (ii) or Theorem \ref{theo211} (ii).
\medskip

Formally, the Hamiltonian of a system with   point interactions ($\delta$-potential) will be written as
$H = T + \sum_{j=1}^n c_j \delta(x_j)$. In order to give a proper definition of $H$ as a \pip\ operator, we rewrite this in momentum representation as
\be\label{eq-ham}
H = T - \sum_{j=1}^n c_j |e^{x_j}_{\nu} \rangle \langle e^{x_j}_{\nu} | ,
\en
where the ``potential" term is a dyadic operator in {Op}$(V_T)$ (see Sec. \ref{subsec:oper}). The exponential functions, which correspond to $\delta$ functions in position representation, are
$$
e^{x}_{\nu}(p)= (2\pi)^{-\nu/2} e^{ix\cdot p}, \, x,p\in \RN^\nu.
$$
The result depends on the dimension $\nu$. Indeed one verifies immediately that
$$
e^x_{1}  \in \H_{\ov 1}(\RN),\; e^x_{\nu} \in \H_{\ov 2}(\RN^\nu)\setminus \H_{\ov 1}(\RN^\nu)\mbox{ for } \nu=2,3; e^x_{\nu}   \not\in \H_{\ov 2}(\RN^\nu) \mbox{ for } \nu\geq 4.
$$
More generally,   given   the set  $\bPhi = \{f _{1},\ldots, f_{n}\}$ of vectors from $ \H_{\ov 2}$
and an arbitrary $n\times n$ matrix $\bB = [B_{ij}]$, one defines the operator
\be\label{eq:dyadicpert}
|\bPhi \rangle \bB \langle \bPhi | := \sum_{i,j=1}^n B_{ij}|f_{i} \rangle \langle f_{j}  |.
\en
Using this notation, we can define the Hamiltonian $H$ as $T$ perturbed by a dyadic of the form \eqref{eq:dyadicpert}.
For the case of point interactions, one takes, of course, $f_{j} = e^{x_j}_{\nu}$.

Using this language, the following results are given in \cite{ghm}. The first result covers the case of a mildly singular perturbation, that is, $f_{k}\in \H_{\ov 1}, \, k = 1, \ldots, n$.

\beprop\label{prop:mildpert}
Let $\bPhi = \{f _{1},\ldots, f _{n}\}, f_{k}\in \H_{\ov 1}$, and let $\bB$ be an invertible $n\times n$ matrix.
Then the natural restriction  of $H = T - |\bPhi \rangle \bB \langle \bPhi | $ is a closed operator in $\H_{0}$. The resolvent of this operator is
$$
 R_ {\lambda}(H) =  R_ {\lambda}(T) -  R_ {\lambda}(T) |\bPhi \rangle \Gamma(\lambda)^{-1}\langle \bPhi |  R_ {\lambda}(T),
$$
where $ R_ {\lambda}(T)$ is the resolvent of $T$ and  $\Gamma(\lambda)= \ip{\bPhi}{ R_ {\lambda}(T) \bPhi} - \bB ^{-1}$.
If $\bB $ is Hermitian, then $H $ is self-adjoint. The points in $\sigma(H)$ that do not belong to $\sigma(T) = [0,\infty)$, i.e., the eigenvalues of $H$ (bound states)  are the solutions of the equation $\det \Gamma(\lambda) =0$. There are at most $n$ real such points.
In addition, $\Gamma(\lambda)$
 can be continued analytically to a Riemann surface and the zeros of this extension will give resonances.

\enprop
By `natural restriction', we mean, of course, restriction in the sense of \pip\ operators, as in Proposition \ref{prop31}. The case of the $\delta$-potentials corresponds to $\bB_{ij} = \delta_{ij}c_j \in \RN$ and $f _j = e^{x_j}_{\nu}$. In the one center case, $n=1$,  the Hamiltonian is $H= T - c \, |e^{x}_{\nu}\rangle\langle e^{x}_{\nu} |$, with resolvent
$$
 R_ {\lambda}(H) =  R_ {\lambda}(T) -  \Gamma(\lambda)^{-1} \, R_ {\lambda}(T)   |e^{x}_{\nu}\rangle\langle e^{x}_{\nu} |  R_ {\lambda}(T).
$$
 From this formula, one can deduce all the spectral properties of $H$. For instance, in one dimension, with $t(p) = p^2$,
  one has  $\Gamma(\lambda)= \ip{e^x_{1}}{ R_ {\lambda}(T) e^x_{1}} -c ^{-1}=  {1}/{2\kappa} - \alpha$, where  $\alpha= c^{-1}$ and
$\kappa^2 = -\lambda, \, \mathrm{Re}\,\kappa \geq 0$.
Thus, for $\alpha> 0$, $H$ has a bound state for $ \kappa_b = 1/2\alpha$, i.e., energy  $E_b \equiv \lambda_b = -1/4 \alpha^2$. As $\alpha \to 0, \, \kappa_b$ tends to $\infty$ and turns into a resonance when $\alpha< 0$.

Whenever at least  one $f_{k} \in \H_{\ov 2}\setminus \H_{\ov 1}$, a case called `strongly singular', the restriction of
$H = T - |\bPhi \rangle \bB \langle \bPhi | $ is no longer self-adjoint, but it admits a family of $n^2$ self-adjoint extensions.
This is, for instance, the case of point interactions in dimension 3. The details may be found in \cite[Sec.7.1.3]{pip-book} and in the original paper \cite{ghm}.
\enex

\beex  {\bf Multiplication by a Dirac $\delta$ in $\cS \subset L^2 \subset \cS\ta$ } \label{ex43}
\\[2mm]
This example is treated in \cite[Ex.5.1]{bello-db-trap}. The operator $M_\delta$ is defined by the relation
$$
\ip{M_\delta f}{g}= \ip{\delta f}{g}= f(0)g(0), \;\forall \, f,g\in \cS(\RN).
$$
Clearly, the only condition to impose on $f$ is that it must be continuous at the origin. Thus the natural LHS for this problem is the scale of Sobolev spaces $V_J = \{W^{k,2}, \, k\in \ZN\}$ \cite[Ex.5.4.21]{pip-book}, with norms
$$
\|f\|_{k,2}=\Big(\int_{\RN} \big(1+|\xi|^2)^{k}|\widehat f(\xi)|^2 \, \ud\xi\Big)^{1/2}
=    \norm{2}{\F^{-1} \big((1+|\cdot|^2)^{k/2}\F (f)\big)}\ .
$$
Note that $V_J \co= \cap_{k\in \ZN} W^{k,2}$ is \emph{not} nuclear, since the embedding of any of the spaces into a bigger one, which is a multiplication operator, cannot be Hilbert-Schmidt. However we can complete the scheme with Schwartz spaces and get
$$
\cS \subset V_J \co \subset W^{0,2} = L^2\subset V_J \subset \cS\ta\, ,
$$
which is the standard RHS used for analyzing singular operators.

In the context of $V_J$, the analysis of  \cite{bello-db-trap} shows that the operator $M_\delta$ is symmetric, has 0 as unique eigenvalue and $J$-spectrum $\sigma^J(M_\delta)= \CN$.
\enex

\beex {\bf  Multiplication by an increasing function in $L^2(\RN)$}   \label{ex44}
\\[2mm]
Taking again the Schwartz RHS $\cS \subset L^2 \subset \cS\ta$, we consider the operator of multiplication $M_\Phi$ by a tempered distribution $\Phi$. This case has been treated in \cite{bello-db-trap}, but only when $\Phi$ is given by a measurable, slowly increasing real function $h$, using again the Sobolev scale. Thus $M_h$ is defined by
$$
\ip{M_h f}{g} = \ip{h}{\ov f g} = \int_\RN h(x) f(x) \ov{g(x)} \ud x, \; f, g \in \cS.
$$
First, $\lambda\in\RN$ can be a genuine eigenvalue of $M_h$ only if $h(x) = a$ a.e.,  for some $a\in\RN$, and then
$\sigma(M_h) = \sigma_p(M_h) = \{a\}$.
Next, the resolvent of $M_h$ is the operator of multiplication by $g=(h-\lambda)^{-1}$.

Assume first that $h$ is bounded. Then, coming back to the
Sobolev scale $V_J = \{W^{k,2}, \, k\in \ZN\}$, the analysis of
\cite{bello-db-trap} shows that  $\rho^{(k,m)}(M_h) = \emptyset, \; \forall\, k,m\in \ZN$,  except for
$\rho^{(0,0)}(M_h) = \CN \setminus \ov{h(\RN)}$, where $\ov{h(\RN)}$ is the closure of the essential range of $h$.
Thus $\rho^{J}(M_h) = \CN \setminus \ov{h(\RN)}$ and $\sigma^{J}(M_h) = \sigma(M_h)= \sigma_c(M_h)=
\sigma_\mathrm{ext}(M_h)= \ov{h(\RN)}$.

On the contrary, if $h$ is slowly increasing and unbounded, such as $h(x) = x$, then again $\rho^{(k,m)}(M_h) = \emptyset, \; \forall\, k,m\in \ZN$, without exception, so that
$\rho^{J}(M_h)  = \emptyset$ and $\sigma^{J}(M_h) = \CN$, whereas $\sigma(M_h)= \sigma_c(M_h)=
\sigma_\mathrm{ext}(M_h)= \ov{h(\RN)}$.

Thus, in both cases, we have a tight rigging.
\enex

\beex {\bf  Multipliers in sequence spaces}\label{ex45}
\\[2mm]
Let $\H$ be a separable \hs, with an orthonormal basis $\{e_n, n=0,1,2\ldots\}$.
Then the space $\H$ is unitary equivalent to the space $\ell^2$ of square integrable sequences, with the usual inner product, via the representation
$f = \sum_{n=1}^\infty  f_n e_n$. Assume now there is a LHS $\{\H_n, n\in\ZN\}$ with central \hs\ $\H_0 = \H$. Correspondingly, we get a LHS $\{\h_n, n\in\ZN\}$ of sequence spaces around $\ell^2$. The typical example is the Schwartz RHS $\cS \subset L^2 \subset \cS\ta$, unitary equivalent to the  RHS of sequences $s\subset \ell^2 \subset s\ta$ via the basis of Hermite functions. In order to introduce a LHS interpolating between $s$ and $s\ta$, consider the \hs s $s_m$ defined as follows, for every $m\in\ZN$:
\be\label{seqspaces}
(f_n) \in s_m \; \Longleftrightarrow\; \sum_{n=1}^\infty  |f_n|^2 (n+1)^m <\infty.
\en
Then $s_m $ is a space of decreasing sequences if $m>0$ and a space of slowly increasing sequences if $m<0$, the duality reads
$(s_m)\ta =  s_{\ov m} $ and one has, as announced,
\be\label{schwspaces}
s= \bigcap_{m\in\ZN} s_m  \quad \mbox{and} \quad s\ta= \bigcup_{m\in\ZN} s_m .
\en

Now consider in $s\ta$ a multiplier \cite{balazs}, that is, an operator $A^{(k)}$ given by $(A^{(k)}c)_n = a_n\, c_n, \, c=(c_n)$,
where the sequence $(a_n)$ satisfies the conditions
\be\label{mult}
|a_n| > 0, \; \forall\, n, \quad \mbox{and} \quad |a_n| \leq (n+1)^{k/2}, \, \forall \, n \mbox{ and some } k\in\ZN.
\en
It follows that $A^{(k)}$ maps $s_m$ continuously into $s_{m-k}$, for every $m\in\ZN$. If $k \gg 1, A^{(k)}$ is a very singular operator.
Let now $m>0$ and $k=2m$. Then $A^{(2m)}$ maps $s_{m}$ into $s_{\ov m}$, so that Theorem \ref{theo35} applies.
If every $a_n$ is real and positive, $A^{(2m)}$ is positive and symmetric in the scale $V_J = \{s_m, \, m\in\ZN\}$, every vector $e_j$ in the canonical basis of $\ell^2$ is an eigenvector of  $A^{(2m)}$, with eigenvalue $a_j$. Thus $\sigma(A^{(2m)})= \sigma_p(A^{(2m)})=
\sigma_\mathrm{ext}(A^{(2m)})= \ov{\{a_j, \, j\in\NN\}}$. In other words, we have a tight rigging.

The preceding example can be generalized to arbitrary weighted sequences, following the discussion in \cite{ant-bal-semiframes1} and in particular
\cite[Sect.3.3]{ant-bal-semiframes2}.  Given   an orthonormal basis $(e_n), n\in \NN,$   in $\H$, define the sequences
 $(\psi_{n}), (\phi_{n})$, with
$\psi_{n}:=m_n^{-1} e_n$,  $\phi_{n}:= {m_n} e_n$,  where $m^{-1} = (m_n^{-1} ) \in\ell^\infty$  has a subsequence converging to zero and $m_n\neq 0, \,\forall \,n$.
Hence the former is an upper semi-frame and not a frame, that is, it satisfies the upper  frame bound, but  not the lower one:
\be
0 < \sum_{n\in \NN} \left| \ip {f} {\psi_n}\right|^2  \le {\rm\sf M}  \norm{}{f}^2 ,  \forall \, f \in \H, \, f\neq 0.
\label{eq:discr-unbddframe}
\vspace{-1mm}\end{equation}
The frame operator associated to the sequence $(\psi_n)$,
defined by
$$
Sf = \sum_{n\in \NN}  \ip {f} {\psi_n}\psi_n,
$$
 is  diagonal, namely,  $S= \mathrm{diag}(m_{n}^{-2})$. Thus $S^{-1}= \mathrm{diag}(m_{n}^{2})$, which is clearly unbounded, and $\psi_{n}  = S^{-1}\phi_{n}$. Considering the scale built on the powers of $S^{-1/2}$, one gets for the central triplet
$$
\H_1 \subset \H_0 = \ell^2  \subset \H_{\ov 1}.
\vspace{-1mm}$$
 The  norm  of $\H_k,  k= 1, 0, \ov{1} $, reads as:
 $$
 \norm{k}{f}^2 = \sum_{n\in \NN} m_{n}^{2k} |f_n|^2, \; k= 1, 0, \ov{1}.
 $$
Next one can consider the full scale $\{\H_j, j\in\ZN \}$ associated to $S^{-1/2}$ and try to identify the end spaces.
For instance, if the sequence $(m_n)$ grows polynomially, one gets the same result: the end spaces $\H_{\infty}(S^{-1/2})=\bigcap_{j} \H_j$,
 resp. $\H_{-\infty}(S^{-1/2}) =\bigcup_{j} \H_{j}$,  still coincide with $s$ and $ s^\times$, respectively.

In that more general context,  multipliers can be defined exactly as in the Schwartz case, with similar results.
Let again $A^{(k)} : (f_n) \mapsto  (a_n\, f_n)$, with
$$
|a_n| > 0, \; \forall\, n, \quad \mbox{and} \quad |a_n| \leq c\, (m_{n})^{k}, \, \forall \, n \mbox{ and some } k\in\ZN.
$$
Then $A^{(k)}$ maps $\H_j$ continuously into $\H_{m-k}$, for every $j\in\ZN$.
Let now $r>0$ and $k=2r$. Then $A^{(2r)}$ maps $\H_{r}$ into $\H_{\ov r}$, so that Theorem \ref{theo35} applies.
If every $a_n$ is real and positive, $A^{(2r)}$ is positive and symmetric in the scale $V_J = \{\H_j, \, j\in\ZN\}$, every vector $e_n$ in the canonical basis of
$\ell^2$ is an eigenvector of  $A^{(2r)}$, with eigenvalue $a_n$. Thus $\sigma(A^{(2r)})= \sigma_p(A^{(2r)})=
\sigma_\mathrm{ext}(A^{(2r)})= \ov{\{a_n, \,n\in\NN\}}$. In other words, we have again a tight rigging.

One may note that the operator
 $S^{-1/2} : \H_j \to \H_{j-1}$ itself is a (mild) multiplier. Even $S^{-1} : \H_1 \to \H_{\ov1}$ fits in the scheme above.
\enex

Another generalization consists in starting from a generalized Riesz basis, in the sense of \cite{bello-trap}, instead of an orthonormal basis. Take again a triplet of \hs s $\H_1 \subset \H_0   \subset \H_{\ov 1}$, with $T: \H_1 \to\H_0 $ a continuous, invertible operator, with bounded inverse  $T^{-1}: \H_0 \to\H_1 $. Then $\{f_n\}$ is a (tight) generalized Riesz basis for $\H_1$ if  $\{Tf_n\}$ is an  orthonormal basis for $\H_0$. It follows that  $\{f_n\}$ is  an  orthonormal basis for $\H_1$ and the norm of the latter is $\norm{1}{f} = \norm{0}{Tf}$. Since $T$ is an unbounded operator in $\H_0 $, with dense domain  $\H_1$,
we may again consider the scale built on the powers of $T$, namely, $\{\H_j, j\in\ZN \}$, where $\H_j= D(T^j)$ with norm  $\norm{j}{f} = \norm{0}{T^jf}$.
In that case, multipliers can be defined as before. In particular, the operator $R^\alpha$ of \cite[Sec.4]{bello-trap}, defined for $f\in \H_1$ by
$$
R^\alpha f =  \sum_{n\in \NN} \alpha_n  \ip {f} {\psi_n}\psi_n, \;  \alpha =(\alpha_n), \, \alpha_n \in \CN,
$$
is such a multiplier.

\beex {\bf  Multipliers in continuous (semi-)frames}\label{ex46}
\\[2mm]
A  construction corresponding to that of Schwartz sequence spaces may be made around $L^2 $ by considering the Hilbert scale built on the powers of
$H_\mathrm{osc}:= -\ud^2/\ud x^2 + x^2$, namely the Hamiltonian of the quantum harmonic oscillator. In that case one interpolates between $\cS$ and
$\cS\ta$ by a scale of \hs s, which are essentially Sobolev-type spaces.

A different approach consists in generalizing the semi-frame context to continuous upper semi-frames, following \cite[Sect.5.2]{ant-bal-semiframes2},
or that of multipliers for continuous frames \cite{bayer}. Let us summarize the first method.

 Let $\H$ be a  Hilbert space and   $X$ a  locally compact space with measure $\nu$. Then a \emph{continuous upper semi-frame} for $\H$  is a family of vectors
   $\Psi:=\{\psi_{x},\, x\in X\}, \,\psi_{x}\in \H $,    indexed by points of $X$,
such that the map  $x \mapsto \ip{f}{\psi_{x}}$  is measurable,  $\forall\,f \in \H$, and
 there exists ${\rm\sf M}<\infty$ such that
\be\label{eq:upframe}
0 < \int_{X}  |\ip{f}{\psi_{x}}| ^2 \, \ud \nu(x)   \leq {\rm\sf M}  \norm{}{f}^2 , \; \forall \, f \in \H, \, f\neq 0 .
\en
Define the   {analysis operator} by the (coherent state) map $C_{\Psi}: \H \to L^{2}(X, \ud\nu)$ given by
$$
(C_{\Psi}f)(x) =\ip{f}{\psi_{x}} , \; f \in \H,
$$
with  adjoint $C_{\Psi}^\ast: L^{2}(X, \ud\nu) \to \H$, called the  {synthesis operator}.
Then the frame operator is $ S:=C_{\Psi}^* C_{\Psi}$ and $ \|C_{\Psi} f\|^2_{L^{2}(X)}= \| S^{1/2}f\|_{\H}^2 = \ip{Sf}{f}$.
 Furthermore,  $C_{\Psi}$ is injective, by virtue of the lower bound,  so that $ C_{\Psi}^{-1} : \mathrm{Ran} (C_{\Psi}) \to \H$ is well-defined.
Thus,  $\Psi$ is a total set in $\H$, the operators $C_{\Psi}$ and $S$ are bounded, $S$ is injective and self-adjoint.
Therefore $ \mathrm{Ran} (S)$ is dense in $\H$, $S^{-1}$ is also self-adjoint,
but unbounded, with dense domain $D(S^{-1}) = \mathrm{Ran} (S)$.

Consider now the operators $G: = C_{\Psi} \,S \,C_{\Psi}^{-1} = C_{\Psi}C_{\Psi}^*$
and $G^{-1}:=C_{\Psi} \,S^{-1} \,C_{\Psi}^{-1} =  {C_{\Psi}^*}^{-1}C_{\Psi}^{-1}$,
 both acting in the Hilbert space $ \ov{\mathrm{Ran} (C_{\Psi})}$, the closure of $\mathrm{Ran} (C_{\Psi})$
in  $L^{2}(X, \ud\nu) $.
Both operators are self-adjoint and positive,
$G$ is bounded and  $G^{-1}$ is densely defined   in $\ov{\mathrm{Ran} (C_{\Psi})}$. Furthermore, they are are inverse of each other on the appropriate domains.

Next one shows  \cite{jpa-sqintegI} that  $\mathrm{Ran} (C_{\Psi})$ is complete in the norm
$$
\norm{\Psi}{F}^2 := \ip{GF}{F}_{L^{2}(X)} = \norm{L^{2}(X)}{C_{\Psi}^{-1}F}^2,
$$
hence it is a \hs, denoted by ${\h}_{\Psi}$, and the map    $C_{\Psi}: {\H} \to {\h}_{\Psi}$ is unitary.
Moreover, since the spectrum of  $G^{-1}$ is bounded away from zero,
the norm $\norm{\Psi}{\cdot}$ is equivalent to the graph norm of {$G^{-1/2} = {\left( G^{-1} \right)}^{1/2}$}.

Proceeding as in the discrete case, one obtains, with continuous and dense range embeddings,
\be
 {\h}_{\Psi} \;\subset\; {\h}_{0}\;\subset \; {\h}_{\Psi}^{\times},
\label{eq:conttriplet}
\end{equation}
where
\begin{itemize}
\item[{\bf .}]  $ {\h}_{\Psi} = \mathrm{Ran} (C_{\Psi})$, which is a Hilbert space for the norm
$\norm{\Psi}{\cdot}=\ip { G^{-1}  \cdot\,}{\, \cdot }^{1/2}_{L^2}$;
\item[{\bf .}]   ${\h}_{0} =\ov{{\h}_{\Psi}} = \ov{ \mathrm{Ran} (C_{\Psi})}$ is the closure of ${\h}_{\Psi}$ in $L^{2}(X, \ud\nu)$;
\item[{\bf .}]
${\h}_{\Psi}^{\times}$ is  the completion of  ${\h}_{0}$  (or ${\h}_{\Psi}$) in the norm
$\norm{\Psi^{\times}}{\cdot}:=\ip {G\cdot\,}{\,\cdot }^{1/2}_{L^2}$,
and the conjugate dual of ${\h}_{\Psi}$.
\end{itemize}
In particular, \eqref{eq:conttriplet} is the central triplet of the scale of Hilbert spaces generated by the powers of $G^{-1/2}$,
namely,  $\h_{j}:= D(G^{-j/2}), jÊ\in \ZN$.
 A concrete example,  given originally in   \cite[Sect.5.2]{ant-bal-semiframes2}, is summarized below. In this approach, multipliers can be defined exactly as in the discrete case.

 Note that, if $\Psi$ is a frame,  both $S^{-1}$ and $ G^{-1}$ are bounded and the three spaces in \eqref{eq:conttriplet} coincide, with equivalent norms, and therefore the scale collapses. This is the case, for instance, in the standard examples such as Gabor (or STFT) multipliers or wavelet multipliers (which are then called Calder\' on-Toeplitz operators). Although such examples are too nice for the present considerations, they have led to a considerable field of research, under the names of time-frequency localization operators or anti-Wick operators. We refer to \cite{bayer} for a comprehensive list of references.

 However, we prefer to follow the formulation of \cite{bayer}. Let $\Psi$ and $\Phi$ be two upper semi-frames (called Bessel mappings in \cite{bayer}) and $m : X\to\CN$ a measurable function.
 Then the operator ${\bf M}_{m,\Psi,\Phi} : \H \to \H$ defined in the weak sense by
 \be\label{multiplier}
 {\bf M}_{m,\Psi,\Phi} f := \int_X m(x) \ip{f}{\psi_x} {\phi_x}\, \ud \nu(x) ,
 \en
is called a continuous Bessel multiplier for $\Psi$ and $\Phi$, with symbol $m$. If  $m\in L^{\infty}(X, \ud\nu)$, the corresponding multiplier is a bounded operator. Clearly one has
$$
{\bf M}_{m,\Psi,\Phi} = C_\Phi^* \, M_m \,C_{\Psi},
$$
where $M_m $ is the operator of multiplication by $m$ in $L^{2}(X, \ud\nu)$. It follows that $({\bf M}_{m,\Psi,\Phi})^* = {\bf M}_{m,\Phi,\Psi}$.
Hence we will consider only the case  $\Phi=\Psi$, so that the corresponding multiplier ${\bf M}_{m,\Psi}:= {\bf M}_{m,\Psi,\Psi}$ is a symmetric operator in the scale defined by $\Psi$.

For $m$ bounded, the operator ${\bf M}_{m,\Psi}$ is bounded, but cannot be compact. In order to generate a compact multiplier, the symbol $m$ must be bounded and vanishing at infinity and, in addition, $\Psi$ must be norm bounded, i.e., $\norm{}{\psi_x}\leq M$, for some constant $M>0$ and almost every $x\in X$. Furthermore, under the same conditions, $m\in L^{p}(X, \ud\nu), 1<p<\infty,$ implies that ${\bf M}_{m,\Psi}$ belongs to the Schatten class
$\C^p$. Thus, in order to obtain a singular operator, unbounded or worse, we have to take for $m$ an unbounded function or  even a distribution. But then we are back to Example  \ref{ex44}, albeit in a general Hilbert scale, generated by an arbitrary upper semi-frame.

To get an example, consider the upper semi-frame $\Psi$ introduced in \cite{ant-bal-semiframes1,ant-bal-semiframes2}, which consists of affine coherent states. Here the Hilbert space is $\H^{(n)}:=L^{2}({\RN}^{+}, r^{n-1}\ud r), n = \hbox{integer} \geq 1 $. The vectors $\psi_x$ are indexed by $x\in \RN$ and are given by
$$
\psi_x(r) = e^{-ixr}\, \psi(r), \quad r\in \RN^+,
$$
where   $\psi$ is admissible if it satisfies  the two conditions
\vspace{-2mm} \begin{align*}
  (i) \;  & \sup_{r \in {\RN}^{+}}{\mathfrak s}(r) = 1 , \mbox {where } {\mathfrak s}(r):=2\pi r^{n-1}|\psi (r)|^{2}  \label{eq518}
\\
 (ii) \; & |\psi (r)|^{2} \neq 0, \; \hbox{except perhaps  at isolated points} \; r \in {\RN}^{+}\nn.
\end{align*}
The frame operator $S$ and its inverse $S^{-1}$ are multiplication operators on $\H^{(n)}$, namely
$$
(S^{\pm1}f)(r) =  [{\mathfrak s}(r)]^{\pm1}f(r).
$$
Since ${\mathfrak s}(r)\leq 1$, the inverse $S^{-1}$ is indeed unbounded and no frame vector $\psi_{x}$ belongs to its domain.
Thus the scale generated by $S^{-1/2}$ consists of the spaces $\H_k = D(S^{-k/2}), k\in \ZN$, with squared norms
$$
 \norm{k}{f}^2= \ip{S^{-k} f} {f}_{\H^{(n)}} =\int_0^\infty |f(r)|^2\, [{\mathfrak s}(r)]^{-k}  \, {r^{n-1} \ud r } .
$$
In the same way, one has
$$
(G^{\pm1}F)(x) = \frac{1}{\sqrt{2\pi}}\int_{{\RN}^{+}} e^{ixr}\, \widehat F(r)\, {[{\mathfrak s}(r)]^{\pm1} \,\ud r, }
$$
and, for every $j\in \ZN, \, j\neq 0,$
$$
(G^{j}F)(x) =  \frac{1}{\sqrt{2\pi}} \int_{{\RN}^{+}}   \widehat F(r)\,{[{\mathfrak s}(r)]^j} \, \ud r.
$$
Accordingly, the associated Hilbert scale consists of the spaces $\h_j = D(G^{-j/2}), j\in \ZN$, with squared norm
$$
 \norm{j}{F}^2  = \ip{G^{-j} F}{F}_{L^2}  =    \int_0^\infty  e^{ixr}\,  |\widehat F(r)|^2\, [{\mathfrak s}(r)]^{-j} \,\ud r  , \;  j\in \ZN, \, j\neq 0.
$$
However, the end spaces of either scale do not seem to have an easy interpretation.

On the scale $\{\H_k , k\in \ZN\}$, a multiplier reads as
 \begin{align*}
 {\bf M}_{m,\Phi} f (r) &= \int_\RN m(x) \ip{f}{\psi_x}\, \psi_x(r) \ud  x,
 \\
 &= \psi(r)  \int_\RN e^{-ixr} m(x) \ip{f}{\psi_x}  \ud  x
\\
 &= \psi(r) \int_\RN e^{-ixr} m(x) \int_0^\infty f(s) \, e^{ixs} \,\ov{\psi(s)}\, s^{n-1} \ud s
\\
 &= \sqrt{2\pi}\,\psi(r) \int_0^\infty f(s)\,\ov{\psi(s)}\,  \widehat m(r-s) \, s^{n-1}\ud s ,
 \end{align*}
where we have freely interchanged the integrals and $ \widehat m$ is the Fourier transform of $m$.

Take for instance $m= \delta$, so that $ \widehat m (s) = 1/\sqrt{2\pi}$. Then ${\bf M}_{m,\Phi} f (r) = \psi(r) \ip{f}{\psi}$, that is,
${\bf M}_{\delta,\Phi}$ is simply the orthogonal projection on $\psi$ in $L^{2}({\RN}^{+}, r^{n-1}\ud r)$.
\enex

\appendix
\section{Partial inner product spaces}

 \subsection{PIP-spaces    and   indexed  PIP-spaces}
\label{subsec-pip}

For the convenience of the reader, we have collected here the main features of partial inner product spaces and operators on them, keeping only what is needed for reading the paper.
Further information may be found in our monograph \cite{pip-book} or our review paper \cite{at-AMP}.

The general framework is that of a \pip\ $V$,  corresponding to the linear compatibility $\com$, that is,
a symmetric binary relation $f \com g$  which preserves linearity.
We   call \emph{assaying subspace} of $V$ a  subspace $S$ such that $S^{\#\#} = S$ and
we denote by ${\F}(V,\com)$   the family of all assaying subspaces of $V$, ordered  by inclusion.
The assaying subspaces are denoted by $V_{r}, V_{q} , \ldots $ and the index set is $F$.  By definition, $q \leq r$ if and only if $V_{q} \subseteq V_{r}$.
 Thus we may write
\begin{equation}
\label{eq:gener2}
f\com g \; \Leftrightarrow \; \exists \;r \in F  \mbox{ such that } f \in V_{r}, g \in V_{\overline{r}}\,.
\en

General considerations \cite{birkhoff}  imply that the family   ${\F}(V,\com):= \{ V_r, r\in {F} \}$, ordered  by
inclusion, is  a complete involutive lattice, {i.e.}, it is stable under the following operations, arbitrarily iterated:
\medskip

\begin{tabular}{lccl}
. involution: &$V_r$                     & $\!\!\!\leftrightarrow\!\!\!$& $V_{\ov{r}}=(V_r)^{\#},$\\
. infimum:   & $V_{p \wedge q}$  &$\!\!\! :=\!\!\!$                & $V_p \wedge V_q = V_p \cap V_q,$   \qquad $(p,q,r \in {F})$\\
. supremum: & $V_{p \vee q}$     &$\!\!\! :=\!\!\!$               &$ V_p \vee V_q = (V_p + V_q)^{\#\#}$.
\end{tabular}
\\[2mm]
\noi The smallest element  of $ {\F }(V,\com)$ is  $V\co = \bigcap_r V_r $ and the greatest element   is $V = \bigcup_r V_r$.

By definition, the index set  ${F}$    is  also a complete involutive  lattice; for instance,
 $$
 (V_{p \wedge q})\co = V_{\ov{p \wedge q}}
=  V_{\ov {p} \vee \ov {q}} = V_{\ov {p}} \vee V_{\ov {q}}.
$$

Given a vector space $V$ equipped with a linear compatibility $\com $, a \emph{partial inner product}   on   $(V, \,{\com})$ is a
   Hermitian form  $\ip{\cdot}{\cdot}$ defined exactly on compatible pairs of vectors.
A \emph{partial inner product space}  (\pip)  is a  vector space $V$ equipped with a linear compatibility  and a partial inner product.

From now on, we will assume that our \pip\  $(V, \com, \ip{\cdot}{\cdot})$ is \emph{nondegenerate},
that is,    $\ip{f}{g} = 0   $ for all $ f \in  V^{\#} $ implies $ g = 0$.  As a consequence,  $(V\co, V)$ and
 every couple $(V_r , V_{\ov r} ), \,  r\in {F}, $  are a  dual pair in the sense of topological vector spaces \cite{kothe}.
Next we assume that every $V_{r}$ carries  its Mackey topology $\tau(V_{r},V_{\ov{r}})$, so that its conjugate dual is $(V_r)^\times = V_{\ov {r}}, \; \forall\, r\in {F} $.
Then,   $r<s$ implies $V_r \subset V_s$, and the embedding operator $E_{sr}: V_r \to V_s$  is continuous and has dense range. In particular, $V\co$ is dense in every $V_{r}$.
 In the sequel, we also assume the partial inner product to be positive definite, $\ip{f}{f}>0$ whenever $f\neq0$.

In fact,   the whole structure can be reconstructed from a fairly
 small subset of $\F$, namely, a \emph{generating}    involutive sublattice $\J$   of $\F(V, \com)$, indexed by $J$, which means that
\be\label{eq:gener}
f\com g \; \Leftrightarrow \; \exists \;r \in J \mbox{ such that } f \in V_{r}, g \in V_{\overline{r}}\,.
\en
The resulting structure is called  an \emph{\ipip} and denoted simply by $V_{J} := (V, \J, \ip{\cdot}{\cdot}) $.

For practical applications, it is essentially sufficient to restrict oneself to the case of an \ipip\ satisfying the following conditions:
\bei
\item [(i)]
 every $V_{r}, r\in J$, is a \hs\ or a reflexive Banach space, so that the Mackey topology $\tau(V_{r},V_{\ov{r}})$ coincides with the norm topology;

\item [(ii)]   there is a unique self-dual, Hilbert,  assaying subspace $V_{o} =V_{\overline{o}}$.

\item [(iii)]  for every $V_r\in\J$, the norm $\|\cdot\|_{\ov{r}}$ on $V_{\ov{r}}=V_{r}^\times$ is the conjugate of the
norm $\|\cdot\|_{r}$ on $V_{r}$. In particular, the partial inner product $\ip{\cdot}{\cdot}$ coincides with the inner product of $V_o$ on the latter.

\eni

\noi In that case, the \emph{\ipip}  $V_{J} := (V, \J, \ip{\cdot}{\cdot}) $ is  called, respectively,
a  \emph{lattice of \hs s} (LHS)  or a  \emph{lattice of Banach spaces} (LBS).
This implies, in addition,  that, for a LHS:
\bei
\item [(i)] for every pair $V_{p}, V_{q}\in \J$, the norm on $V_{p\wedge q}:=V_{p}\cap V_{q}$ is equivalent to the projective norm, given by
\be
\| f \|_{p\wedge q}^2 = \| f \|_{p}^2  + \| f \|_{q}^2, \, \mbox{}
\label{projnorm}
\en
\item [(ii)] for every pair $V_{p}, V_{q}\in \J$, the norm   on  $V_{p\vee q} := V_{p} + V_{q}$, the vector sum, is equivalent to the inductive norm
\be
\| f \|_{p\vee q}^2 =
\inf_{f=g+h}\left(\| g \|_{p}^2  + \| h \|_{q}^2\right), \; g \in V_{p}, f \in V_{q}\, .
\label{indnorm}
\en
\eni
Similar formulas are used in the LBS case, simply omitting the squares. These norms come from interpolation theory \cite{berghlof}.

Note that $V\co, V $ themselves usually do \emph{not} belong to the family $\{V_{r}, \,r\in J\}$, but they can be recovered  as
$$
V\co = \bigcap _{r\in J}V_{r}, \quad V =\sum_{r\in J}V_{r}.
$$
 A standard, albeit trivial,  example is that of a Rigged Hilbert space (RHS) $\Phi \subset \H \subset \Phi\co$
(it is trivial because the lattice $\F$ contains only three elements).

Familiar concrete examples are sequence spaces, with $V = \omega$    the space    of \emph{all} complex sequences $x = (x_n)$, and
spaces of locally integrable functions with $V =L^1_{\rm loc}(\RN, \ud x)$, the space of Lebesgue measurable functions, integrable over compact subsets.

\subsection{Operators on \ipip s}
\label{subsec:oper}

Let $V_{J}$   be a nondegenerate \ipip\ (in particular, a LHS or a LBS). Then  an \emph{operator} on $V_J$  is a map
from a subset $\D (A) \subset V$ into $V$, such that
\smallskip

(i) $\D(A) = \bigcup_{q\in {\sf d}(A)} V_q$, where ${\sf d}(A)$ is a nonempty subset of $J$;
\smallskip

(ii)  For every $q \in  {\sf d}(A )$, there exists $p\in J$ such that the restriction of $A$ to $V_{q}$ is a continuous linear map into $V_{p}$ (we denote this restriction by $A_{pq})$;
\smallskip

(iii) $A$ has no proper extension satisfying (i) and (ii).
\medskip

\noi We denote by Op$(V_J,)$  the set of all operators on   $V_J$ .
 The continuous linear operator $A_{pq}: V_q \to V_{p}$ is called a \emph{representative} of $A$.
The properties of $A$ are conveniently described
 by the set ${\sf j}(A)$ of all pairs $ (q,p )\in  J\times J$ such that $A$ maps $V_{q}$ continuously into $V_{p}$
   Thus the operator $A$ may be identified with   the collection of its representatives,
\be\label{eqj(A)}
A \simeq \{ A_{pq}: V_{q} \to V_{p} : (q,p ) \in  {\sf j}(A)\}.
\en
It is important to notice that  an operator is uniquely determined by \emph{any} of its representatives, in virtue  of Property (iii): there are no extensions for \pip¥ operators.

 We will also need the following sets:
\vspace*{-2mm}\begin{align*}
{\sf d}(A) &= \{ q \in J : \mbox{there is a } \,   p \; \mbox{such that}\; A_{pq} \;\mbox{exists} \},
\\
{\sf i}(A) &= \{ p \in J : \mbox{there is a } \, q \; \mbox{such that}\; A_{pq} \;\mbox{exists} \}.
\end{align*}
\\[-4mm]
The following properties are immediate:
\bei
\item [{\bf .}]
${\sf d}(A)$ is an initial subset of $J$:  if $q \in {\sf d}(A)$ and $q' < q$, then $q' \in {\sf d}(A)$, and $A_{pq'} = A_{pq}E_{qq'}$,
 where  $E_{qq'}$ is a representative of the unit operator.

\item [{\bf .}]
${\sf i}(A)$ is a final subset of $J$: if $p \in {\sf i}(A)$ and $p' > p$, then $p' \in {\sf i}(A)$ and $A_{p'q} = E_{p'p} A_{pq}$.
\eni

Although an operator may be identified with a separately continous  sesquilinear form on $V^\# \times V^\#$, or a conjugate linear, continuous map
$V^\#$ into $V$ it is more useful to keep also the \emph{algebraic operations} on operators, namely:
 \bei
\vspace*{-1mm}\item[(i)] \emph{ Adjoint:}
every $A \in\mathrm{Op}(V_J)$ has a {unique} adjoint $A\ta \in \mathrm {Op}V_J)$, defined by
\be\label{eq:adjoint}
\ip {A\ta y} {x} = \ip  {y} { Ax}   , \;\mathrm {for}\,  x \in V_q, \, q\in{\sf d}(A) \;\mathrm {and }\;\, y\in V_{\ov{p}}, \, p \in{\sf i}(A),
\en
that is,    $(A\ta)_{\ov{q}\ov{p}} = (A_{pq})' $, where $(A_{pq})': V_{\ov{p}} \to  V_{\ov{q}}$  is the  adjoint map  of $A_{pq}$.
Furthermore,  one has $A\ta{}\ta = A, $ for every $ A \in {\rm Op}(V_J)$: no extension is allowed, by the maximality condition (iii)  of the definition.

\item[(ii)]
\emph{Partial multiplication:}
 Let $A, B \in   \mathrm{Op}(V_J )$. We say that the product $BA$ is defined if and only  if
there is a $r \in{\sf i}(A) \cap{\sf d}(B)$, that is, if and only if   there is a continuous factorization through some $V_r$:
\be\label{eq:mult}
V_q \; \stackrel{A}{\rightarrow} \; V_r \; \stackrel{B}{\rightarrow} \; V_p , \quad\mbox{{i.e.},} \quad  (BA)_{pq} = B_{pr} A_{rq}, \,\mbox{ for some } \;
q \in{\sf d}(A) , p\in {\sf i}(B).
\en
\eni
{Of particular interest are \emph{symmetric} operators, defined as those operators satisfying the relation
$A\ta  =  A$, since these are the ones that could generate self-adjoint operators in the central \hs, for instance by the celebrated KLMN theorem, suitably generalized to the \pip\ environment \cite[Section 3.3]{pip-book}.}

Concerning the adjoint, we note that  ${\sf j}(A\ta) = {\sf j}\ta(A):=\{(\ov{p},\ov{q}):  (q,p)\in {\sf j}(A)\} \subset J \times J$.
Also,  ${\sf j}(A\ta)$ is obtained by reflecting ${\sf j}(A)$  with respect to the anti-diagonal  $\{(r,\ov{r}), r \in J\}$.
{In particular, if $A$ is symmetric, ${\sf j}(A)$ is symmetric with respect to the antidiagonal. Therefore, if $(r,\ov{r})\in {\sf j}(A)$, then  $(r,\ov{r})\in {\sf j}(A\ta)$ as well.}

For a LBS or a LHS, it turns out that, for any operator $A \in \mathrm{ Op}(V_{J})$, the sets ${\sf d}(A)$ and ${\sf i}(A)$ are both sublattices of $J$. This implies that the domain ${\D}(A)$ of  $A$ is a vector subspace of $V$.
In addition,  according to    \cite[Lemma 3.3.29]{pip-book}, if $(q,p)$ and $(t, s)$ belong to ${\sf j}(A)$, so do
$(q\wedge t,p\wedge s)$ and $(q\vee t,p\vee s)$.
Actually, this property remains true if $V_{J}$ is a \emph{projective} \ipip, that is, for each pair $p,q$, the Mackey topology on
$V_{p \wedge q} = V_{p}  \cap V_{q}$ coincides with the projective topology inherited from $V_{p}$ and $V_{q}$.
Thus Proposition \ref{prop-inverse}  still holds true in that more general case.

Moreover,   $\mathrm{  Op}(V_{J})$ is a (non associative) partial *-algebra, with respect to the  partial multiplication of operators \cite{ait_book}.
For studying the lattice properties of $\mathrm{ Op}(V_{J})$, it is useful to consider the sets
 \be
 \O_{pq} = \{ A \in  \mathrm{Op}(V_{J}): A_{pq} \; \mbox{exists} \}.
\end{equation}
Thus
$$
A\in  \O_{pq}\; \Longleftrightarrow \; (q,p) \in {\sf j}(A).
$$
As compared with the notations of \cite{bello-db-trap},  $A\in  \O_{pq}$ is the equivalent to $A\in \C(\E,\F)$, with the interspaces
$\E=V_q, \F = V_p$.

Particularly useful are the \emph{dyadic operators}, that is, rank 1 operators   of the form $|f \rangle \langle g |, \,f ,g \in
V$, defined as
 $$
 | f \rangle \langle g |\, (h) = \ip{g} {h} f \quad h \in V\co.
 $$
Since our inner product $\ip{\cdot}{\cdot}$ is linear in the \emph{second} factor, we have $ |f \rangle \langle g |
:= f \otimes \ov{g}$.

Of course, this operator may be extended to any $V_r$ such that $g\in V_{\ov r}$. The adjoint of $|f \rangle \langle g|$ is   $|g \rangle \langle f|$. One constructs in the same way operators between different spaces and  finite linear combinations of dyadics.

\end{document}